\documentstyle[epsfig,12pt]{article}
\newlength{\dinwidth}
\newlength{\dinmargin}
\newcommand{\ba}{\begin{array}}
\newcommand{\ea}{\end{array}}
\newcommand{\be}{\begin{equation}}
\newcommand{\ee}{\end{equation}}
\newcommand{\bea}{\begin{eqnarray}}
\newcommand{\eea}{\end{eqnarray}}

\def\parallel{| \hskip-0.03cm |}

\def\bee{\begin{eqnarray}}
\def\eee{\end{eqnarray}}
\def\be{\begin{equation}}
\def\ee{\end{equation}}

\begin{document}
\thispagestyle{empty}
\addtocounter{page}{-1}
\begin{flushright}
SNUTP 98-016\\
{\tt hep-th/9803001}\\
Expanded Version\\
\end{flushright}
\vspace*{1.3cm}
\centerline{\Large \bf Macroscopic Strings as Heavy Quarks}
\vspace*{0.3cm}
\centerline{\Large \bf in}
\vspace*{0.3cm}
\centerline{\Large \bf Large $N$ Gauge Theory and Anti-de Sitter Supergravity
\footnote{
Work supported in part by the KRF International Collaboration Grant, 
the KOSEF Interdisciplinary Research Grant 98-07-02-01-5, and the KOSEF 
Leading Scientist Grant.}}
\vspace*{1.2cm} \centerline{\large \bf Soo-Jong Rey and Jung-Tay Yee}
\vspace*{0.8cm}
\centerline{\it School of Physics \& Center for Theoretical Physics}
\vspace*{0.3cm}
\centerline{\it Seoul National University, Seoul 151-747 \rm KOREA}
\vspace*{1.5cm}
\centerline{\Large\bf abstract}
\vspace*{0.5cm}
We study some aspects of Maldacena's large $N$ correspondence between 
${\cal N}=4$ superconformal gauge theory on D3-brane and maximal supergravity 
on $AdS_5 \times S_5$ by introducing macroscopic strings as heavy 
(anti)-quark probes. The macroscopic strings are semi-infinite Type IIB 
strings ending on D3-brane world-volume. We first study deformation and 
fluctuation of D3-brane when a macroscopic BPS string is attached. We find
that both dynamics and boundary conditions agree with those for macroscopic
string in anti-de Sitter supergravity. As a by-product we clarify how 
Polchinski's Dirichlet and Neumann open string boundary conditions arise
dynamically. We then study non-BPS macroscopic string anti-string pair 
configuration as physical realization of heavy quark Wilson loop. We obtain 
$Q \bar Q$ static potential from the supergravity side and find that the 
potential exhibits nonanalyticity of square-root branch cut in `t Hooft 
coupling parameter. We put forward the nonanalyticity as prediction for 
large-N gauge theory at strong `t Hooft coupling limit. By turning on 
Ramond-Ramond zero-form potential, we also study $\theta$ vacuum angle 
dependence of the static potential. 
We finally discuss possible dynamical realization of heavy $N$-prong 
string junction and of large-N loop equation via local electric field and 
string recoil thereof. Throughout comparisons of the AdS-CFT correspondence, 
we find crucial role played by `geometric duality' between UV and IR scales
on directions perpendicular to D3-brane and parallel ones, explaining how 
$AdS_5$ spacetime geometry emerges out of four-dimensional gauge theory 
at strong coupling. 
\vspace*{1.1cm}

\baselineskip=18pt
\newpage

%%%%%%%%%%%%%%%%%%%%%%%%%%%%%%%%%%%%%%%%%%%%%%%%%%%%%%%%%%%%%%%%%%%%%%%%%%%%%
\section{Introduction}
%%%%%%%%%%%%%%%%%%%%%%%%%%%%%%%%%%%%%%%%%%%%%%%%%%%%%%%%%%%%%%%%%%%%%%%%%%%%%
\setcounter{equation}{0}
With better understanding of D-brane dynamics, new approaches to outstanding
problems in gauge theory have become available. One of such problems is
regarding 
the behavior of $SU(N)$ gauge theory in the large $N$ limit~\cite{thooft}: 
$N \rightarrow \infty$ with `t Hooft coupoing 
$g^2_{\rm eff} = g^2_{\rm YM} N$ fixed. 
Planar diagram dominance as shown first by `t Hooft has been regarded as an 
indicative of certain connection to string theory but it has never been clear 
how and to what extent the string is
related to the fundamental string. 
Recently, built on earlier study of near-horizon
geometry of D- and M-branes~\cite{douglas} 
and their absorption and Hawking emission processes~\cite{klebanov}, 
Maldacena has put forward a remarkable proposal to the large $N$ 
behavior~\cite{maldacena}. According to his proposal, large $N$ limit of 
$d$-dimensional conformal
field theories with sixteen supercharges is governed in dual description 
by maximal supergravity theories (chiral or non-chiral depending on $d$)
with thirty-two supercharges that are compactified on $AdS_{d+1}$ times
internal round sphere. 
Extentions to nonconformally invariant field theories \cite{yank} and new 
results ~\cite{klebanov2, horowitz, ferrara, witten}
extending Maldacena's proposal have been reported.  

The most tractible example of Maldacena's proposal is four-dimensional
${\cal N} = 4$ super Yang-Mills theory with gauge group $SU(N)$. The 
theory is superconformally invariant with vanishing beta function and
is realized as the world-volume theory of $N$ coincident D3-branes of
Type IIB string theory. The latter produces near horizon geometry
of $AdS_5 \times S_5$, where $\lambda_{\rm IIB} = g^2_{\rm YM}$, 
the radius of curvature $\sqrt{g_{\rm eff}} \ell_{\rm s}$ 
and self-dual flux of  $Q_5 = {1 \over 2 \pi} \int_{S_5} H_5 = N$ units. 
By taking $\lambda_{\rm IIB} \rightarrow 0$ while keeping $g_{\rm eff}$ 
large in the large $N$ limit, the classical Type IIB string theory is 
approximated by the compactified supergravity. 

In this paper, we study some aspects of large $N$ behavior of superconformal
$d=4, {\cal N}=4$ Yang-Mills theory with gauge group $SU(N)$ 
from the perspectives of Maldacena's
proposal. In particular, we pay attention to charged particles in the theory.
It is well-known that, conformal invariance imposes vanishing
electric current as an operator equation, leading only to a trivial theory.
It has been argued that~\cite{argyres}, to obtain a nontrivial conformally
invariant fixed point, there must be nonvanishing electric and magnetic 
states in the spectrum. As such, it would be most desirable to investigate 
the theory with charge particles in detail. Massless charged particles, 
even though being of our ultimate interest, would
be rather delicate because their long-range fields are exponentially 
suppressed due to conformal invariance. Thus, in this paper, we would like to 
concentrate exclusively on heavy electric and magnetic particles. 

The idea is very simple. The spectrum
of $d=4, {\cal N}=4$ super-Yang-Mills theory contains BPS spectra carrying
electric and magnetic charges $(p,q)$. Extending the 
Maldacena's conjecture, one expects that the correspondence between gauge 
theory and supergravity continues to hold even when heavy charged particles 
are present. In particular, dynamics of BPS particles should match between 
gauge theory and supergravity descriptions. On the supergravity side, 
charged particle may be described by a macroscopic Type IIB $(p,q)$ string 
that ends on the D3-branes. For example, ending on D3-brane, a macroscopic 
fundamental $(1,0)$ string represents a static, spinless quark transforming 
in the defining representation of the $SU(N)$ gauge group. On the gauge theory 
side, one can also describe the BPS charged particles as worldvolume solitons 
on the D3-brane. Using Born-Infeld worldvolume action, Callan and Maldacena 
\cite{callanmaldacena} have shown that the worldvolume BPS solitons are
identical to the Type IIB $(p,q)$ string ending on the D3-branes. 
Thus, equipped with both supergravity and worldvolume descriptions,
one would be able to test Maldacena's conjecture explicitly even when the
conjecture is extended to include heavy charged states.

Using aforementioned correspondence between heavy charged states and 
macroscopic strings, we will prove that static quark-antiquark potential 
comes out of regularized energy of a static configuration of open Type IIB 
string in anti-de Sitter supergravity background. We will find that the static 
potential is of Coulomb type, the unique functional form consistent with 
the underlying conformal invariance \cite{peskin}, and, quite surprisingly, 
is proportional to the {\sl square-root} of `t Hooft coupling parameter. We
interpret the nonanalyticity as an important prediction of Maldacena's
conjecture on super-Yang-Mills theory in large-$N$, strong `t Hooft coupling 
limit. 

In due course of the study, we will elaborate more on boundary conditions 
that the worldvolume BPS soliton satisfies at the throat. According to 
Polchinski's prescription, open string coordinates in perpendicular and 
parallel directions to D-brane should satisfy Dirichlet and Neumann boundary 
conditions, respectively. For the worldvolume BPS soliton, we will show that 
these boundary conditions arise quite naturally as a consequence of 
self-adjoint extension \cite{simon, cheon} of small fluctuation operators 
along the elongated D3-brane worldvolume of BPS soliton.

This paper is organized as follows. In Section 2, we study dynamics of 
a macroscopic Type IIB string, using the Nambu-Goto formulation, in the 
background of multiple D3-branes. 
In section 3, the result of section 2 is compared with dynamics of Type IIB 
string realized as worldvolume BPS soliton on the D3-brane. We find that the 
two descriptions are in perfect agreement. As a bonus, we will be able to 
provide dynamical account of Polchinski's D-brane boundary conditions 
out of self-adjointness of the low-energy string dynamics.
In section 4, we also study large $N$ resummed Born-Infeld theory and find
D3-brane world-volume soliton that corresponds to semi-infinite string and
to massive charged particle on the D3-brane.  
In section 5, we consider a heavy quark and anti-quark pair configuration, 
again, from both the large-$N$ resummed Born-Infeld and the supergravity sides. 
As a prototype nonperturbative quantity, we derive static inter-quark potential.
Results from both sides are qualitatively in good agreement and, most
significantly, displays surprising nonanalytic behavior with respect to the 
`t Hooft coupling. 
We also point out that the static inter-quark potential suggests a dual 
relation between the ultraviolet (infrared) limit of supergravity side 
and the infrared (ultraviolet) limit of the gauge theory side, which we refer
as UV-IR geometry duality. 
In Section 6, we speculate on possible relevance of conformal invariance to 
the large-N Wilson loop equation and realization of exotic hadron states in large-$N$ gauge theory via $N$-pronged string networks on the supergravity side. 

%%%%%%%%%%%%%%%%%%%%%%%%%%%%%%%%%%%%%%%%%%%%%%%%%%%%%%%%%%%%%%%%%%%%%%%%%%%%
\section{String on D3-Brane: Supergravity Description}
%%%%%%%%%%%%%%%%%%%%%%%%%%%%%%%%%%%%%%%%%%%%%%%%%%%%%%%%%%%%%%%%%%%%%%%%%%%%
Consider $N$ coincident planar D3-branes (thus carrying total  
Ramond-Ramond charge $N \equiv \oint_{S_5}
H_5 = \oint_{S_5} H^*_5$), all located at ${\bf x}_\perp = 0$.  
Supergravity background of the D3-branes is 
given by
\be
ds^2_{\rm D3} = G_{\mu \nu} dx^\mu d x^\nu 
= {1 \over \sqrt G} \left( -dt^2 + d{\bf x}^2_{\parallel} \right)
+ {\sqrt G} \, \left( d r^2 + r^2 d \Omega_5^2 \right),
\label{d3metric}
\ee
where
\be
G (r) = 1 +  g_{\rm eff}^2 \left({{\sqrt \alpha'} \over r}\right)^4. 
\label{G}
\ee
In the strong coupling regime $g_{\rm eff} \rightarrow \infty$,
the geometry described by near horizon region is given by anti-de Sitter 
spacetime $AdS_5$ times round $S_5$. For extremal D3-branes, 
the dilaton field is constant everywhere. As such, up to the string coupling 
factors, the supergravity background Eq.(\ref{d3metric}) coincides with the 
string sigma-model background. 

We would like to study dynamics of a test Type IIB fundamental string that 
ends on the D3-branes~\footnote{By $SL(2,{\bf Z})$ invariance of Type IIB 
string theory, it is straightforward to extend the results to the situation 
where the test string is a dyonic $(p,q)$ string \cite{callanmaldacena}.}. 
Let us denote the string coordinates $X^\mu(\sigma, \tau)$, where 
$\sigma, \tau$ parametrize the string worldsheet.
Low-energy dynamics of the test string may be described in terms of 
the Nambu-Goto action, whose Lagrangian is given by
\be
L_{\rm NG} = T_{(n, 0)} \int d \sigma \sqrt{-{\rm det} \, h_{ab} }
+ L_{\rm boundary},
\ee
where $T_{(n, 0)} = n / 2 \pi \alpha'$ denotes the string tension  
($n$ being the string multiplicity, which equals to the electric charge on the
D3-brane world-volume), $L_{\rm boundary}$ signifies appropriate open string
boundary condition at the location of D3-brane, on which we will discuss 
more later, and $h_{ab}$ is the induced metric of the worldsheet:
\be
h_{ab} = G_{\mu \nu} (X) \partial_a X^\mu \partial_b X^\nu.
\label{inducedmetric}
\ee
For the background metric $G_{\mu \nu}$, our eventual interest is the case 
$g_{\rm eff} \rightarrow \infty$, so that the anti-de Sitter spacetime is 
zoomed in. In our analysis, however, we will retain the asymptotic flat 
region. Quite amusingly, from such an analysis, one will be able to extend 
the Polchinski's description of boundary conditions for an open string ending 
on D-brane in the $g_{\rm st} = 0$ limit, where an exact conformal field 
theory description is valid, to an interacting string ($g_{\rm st} \ne 0$) 
regime. 

To find out relevant string configuration, we take $X^0 = t = \tau$ 
and decompose nine spatial coordinates of the string into:
\be
{\bf X} = {\bf X}_{\parallel} + {\bf X}_\perp.
\label{decompose}
\ee
Here, ${\bf X}_{\parallel}, {\bf X}_\perp$ represent test string coordinates 
longitudinal and transverse to the D3-brane. The transverse coordinates
${\bf X}_\perp$ may be decomposed further into radial coordinate $\alpha'$U 
and angular ones $\bf \Omega_5$. In the background metric Eq.(\ref{d3metric}), 
straightforward calculation yields ( $\dot{} \equiv \partial_t , \,\, 
{}' \equiv \partial_\sigma$ )
\bee
h_{00} &=& {\sqrt G} \, {{\dot {\bf X}}_\perp}^2 - {1 \over \sqrt G}
\,  \left( 1 - {{\dot {\bf X}}_{\parallel}}^2 \right)
\nonumber \\
h_{11} &=& {\sqrt G} \, \left( {{\bf X}'_\perp}^2 \right) 
+{1 \over \sqrt G} \, {{\bf X}_{\parallel}' }^2 
\nonumber \\
h_{01} &=& {1 \over \sqrt G} \, \dot {\bf X}_{\parallel} \cdot 
{\bf X}_{\parallel}'
+ {\sqrt G} \, {\dot {\bf X}}_\perp \cdot {\bf X}'_\perp,
\label{eq9}
\eee
where $G = G(|{\bf X}_\perp|)$.
From this, for a static configuration, is derived the Nambu-Goto Lagrangian:
\be
L_{\rm NG} \rightarrow \int d \sigma \sqrt{ {{\bf X}'_\perp}^2 
+ {1 \over G} {{\bf X}_{\parallel}'}^2 }.
\ee
From the equations of motion:
\bee
\left(  {    {\bf X}'_\perp \over \sqrt{ {{\bf X}'_\perp}^2 + {1 \over G}
{{\bf X}'_{\parallel}}^2 }        }
\right)' &=& {{\bf X}_{\parallel}'}^2 (\nabla_{{\bf x}_\perp} G^{-1})
\nonumber \\
\left( { {1 \over G} {\bf X}'_{\parallel}
\over \sqrt{   {{\bf X}'_\perp}^2 + {1 \over G} { {\bf X}'_{\parallel} }^2  }
}
\right)' &=& 0
\eee
it is easy to see that the solution relevant to our
situation is when ${\bf X}'_{\parallel} = 0$ (a class of solutions with 
${\bf X}'_{\parallel} \ne 0$ corresponds to a string bended along D3-brane, 
some of which will be treated in Section 4).
Solving the equation for ${\bf X}_\perp$, one finds $\sigma = \alpha' {\rm U}$ 
and ${\bf \Omega}_5$ constant.
This yields precisely the static gauge configuration 
\be
X^0 = t = \tau \, \hskip0.75cm \alpha' {\rm U} = r \,\,.
\label{timegauge}
\ee

\subsection{Weak Coupling Limit}
Consider the low-energy dynamics of the test macroscopic string in the
weak coupling regime, $\lambda_{\rm IIB} \rightarrow 0$. In this regime, 
the radial function part in Eq.(2) can be treated perturbatively.  
Expanding the Nambu-Goto Lagrangian around the static gauge configuration,
Eq.(\ref{timegauge}), one derives low-energy effective Lagrangian up to 
quartic order:
\be
L_{\rm NG} = {T_{(n,0)} \over 2} \int_0^\infty \! d r \,
\Big[ \left( {\dot {\bf X}_{\parallel} }^2 - {1 \over G }
\, {{\bf X}_{\parallel}'}^2 \right)
+  \left( G \, {\dot {\bf X}_\perp }^2 - {{\bf X}'_\perp}^2 \right)
+ \left( \dot {\bf X}_{\parallel} \cdot {\bf X}'_\perp
- \dot {\bf X}_\perp \cdot {\bf X}'_{\parallel} \right)^2  \Big] \, .
\label{fluclagrangian}
\ee
At the boundary $r = 0$, where the test string ends on the D3-brane, a 
suitable boundary condition has to be supplemented. The boundary condition
should reflect the fact that the string is attached to the D3-brane 
dynamically and render the fluctuation wave operator self-adjoint.

Let us introduce a tortoise worldsheet coordinate $\sigma$:
\be
{ d r \over d \sigma } = {1 \over {\sqrt G}} \equiv \cos \theta(r);
\hskip0.75cm (-\infty < \sigma < + \infty),
\label{eq11}
\ee
in terms of which the the spacetime metric Eq.(\ref{d3metric}) becomes 
conformally flat:
\be
ds^2_{\rm D3} = {1 \over \sqrt G} \left( -dt^2 + d{\bf x}_{\parallel}^2
+ d \sigma^2 \right) + {\sqrt G} \, r^2 \, d \Omega_5^2.
\label{eq12}
\ee
Quadratic part of the low-energy effective Lagrangian is 
\be
L_{\rm NG} = {T_{(n,0)} \over 2} \int_{-\infty}^{+\infty} d \sigma \, 
\Big[ {1 \over \sqrt G}
\left( (\partial_t {\bf X}_{\parallel})^2 - (\partial_\sigma {\bf X}_{\parallel})^2 \right)
+ 
{\sqrt G} \left( (\partial_t {\bf X}_\perp)^2 - (\partial_\sigma {\bf X}_\perp)^2
\right) 
\, \Big],
\label{eq13}
\ee
which reflects explicitly the conformally flat background Eq.(\ref{eq12}).
The Lagrangian clearly displays the fact
that both parallel and transverse fluctuations
propagate at the speed of light, despite the fact that both mass density 
and tension of the string are varying spatially.

Note that, in the tortoise coordinate Eq.(\ref{eq11}, \ref{eq12}), 
$\sigma \rightarrow - \infty$ corresponds to near D3-brane 
$r \rightarrow 0$, while $\sigma \rightarrow + \infty$ is the asymptotic 
spatial infinity $r \rightarrow \infty$.
In the limit $g_{\rm eff} \rightarrow \infty$, the boundary of anti-de Sitter
spacetime is at $\sigma = 0$. Therefore, to specify dynamics of the open
test string, appropriate self-adjoint boundary conditions has to be 
supplemented at $\sigma = -\infty$ and at $\sigma = 0$ if the anti-de Sitter
spacetime is zoomed in. To analyze the boundary conditions, we now examine 
scattering of low-energy excitations off the D3-brane.

For a monochromatic transverse fluctuation 
${\bf X}_\perp (\sigma, t) = {\bf X}_\perp (\sigma) e^{ - i \omega t}$, 
unitary transformation ${\bf X}_\perp (\sigma) 
\rightarrow G^{-1/4} {\bf Y}_\perp (\sigma) $ 
combined with change of variables
$\sigma \rightarrow \sigma / \omega, r \rightarrow r / \omega, 
g_{\rm eff} \rightarrow g_{\rm eff} / \omega$ where $\epsilon \equiv \sqrt{
g_{\rm eff} } \omega$
yields the fluctuation equation into a one-dimensional Schr\"odinger equation
form:
\be
\left[ - {d^2 \over d \sigma^2} + V_\perp (\sigma) \right] 
{\bf Y}_\perp (\sigma) = +1 \cdot
 {\bf Y}_\perp (\sigma),
\ee
where the analog potential $V(\sigma)$ is given by: 
\bee
V_\perp (\sigma) &=& 
-{1 \over 16} G^{-3} \left[ 5 (\partial_r G)^2 - 4 G (\partial^2_r G) \right]
\nonumber \\
&=& {5 \epsilon^{-2} \over ( r^2/\epsilon^2 + \epsilon^2/r^2)^3}.
\eee
For low-energy scattering, $\epsilon \rightarrow 0$,the potential may be 
approximated by $\delta$-function~\footnote{This is essentially 
the same argument as Callan and Maldacena~\cite{callanmaldacena, leepeetthorlacius}. }.
We now elaborate more for justification of their approximation.
This analog potential has a maximum at $r = \epsilon$. In terms of $\sigma$
coordinates, this is again at $\sigma \approx {\cal O}(\epsilon)$. We thus
find that the one-dimensional Schr\"odinger equation has a delta function-like
potential. For low energy scattering, the delta function gives rise to
Dirichlet boundary condition. An interesting situation is when $g_{\rm eff} 
\rightarrow 0 $. The distance between $r=0$ and $r = \epsilon$ becomes zero. 
Therefore, the low-energy scattering
may be described by a self-adjoint extension of free Laplacian operator
at $r = 0$. 

Similarly, for a monochromatic parallel fluctuation 
${\bf X}_{\parallel} (t, \sigma) = {\bf X}_{\parallel} 
(\sigma) e^{-i \omega t}$, unitary transformation
${\bf X}_{\parallel} = G^{1/4} {\bf Y}_{\parallel}$ combined with the same
change of variables yields:
\be
\left[ - {d^2 \over d \sigma^2} + V_{\parallel} (\sigma) \right]
{\bf Y}_{\parallel}(\sigma)
= + 1 \cdot {\bf Y}_{\parallel} (\sigma),
\ee
where
\bee
V_{\parallel} (\sigma) &=& {1 \over 16} G^{-3} 
\Big[ 7 (\partial_r G)^2 - 4 G (\partial_r^2 G) \Big]
\nonumber \\
&=& - {(5r^2/\epsilon^2 - 2 \epsilon^2/r^2) \over (r^2/\epsilon^2 + 
\epsilon^2/r^2)^3}.
\eee
By a similar reasoning as the transverse fluctuation case,
for low-energy scattering $\epsilon \rightarrow 0$,
it is straightforward to convince oneself that the analog
potential approches $\delta'(\sigma - \epsilon)$ -- derivative of
delta function potential. It is well-known that $\delta'$-potential yields 
Neumann boundary condition~\cite{simon, cheon}. An interesting point is that 
the scattering center is {\sl not} at the brane location $r = 0$ naively 
thought from conformal field theory reasoning but a distance 
${\cal O}(\epsilon)$ away. 

We have thus discovered that the Polchinski's conformal field theoretic
description for boundary conditions of an open string ending on D-branes 
follows quite naturally from dynamical considerations of string fluctuation 
in the low-energy, weak `t Hooft coupling $g_{\rm eff} \rightarrow 0$ 
limit.

%%%%%%%%%%%%%%%%%%%%%%%%%%%%%%%%%%%%%%%%%%%%%%%%%%%%%%%%%%%%%%%%%%%%%%%%%%%%
\subsection{Strong Coupling Limit}
%%%%%%%%%%%%%%%%%%%%%%%%%%%%%%%%%%%%%%%%%%%%%%%%%%%%%%%%%%%%%%%%%%%%%%%%%%%%
Let us now consider the low-energy dynamics of the test string in the strong
coupling regime, $g_{\rm eff} \rightarrow \infty$. Suppose $N$ coincident 
D3-branes are located at $\vert {\bf x}_\perp \vert \equiv \ell_s^2 
{\rm U} = 0$ and, 
in this background, probe D3-brane of charge $k$ ($k \ll N$) is located at 
${\bf x}_\perp = {\bf x}_0$. We will be considering a macroscopic fundamental 
Type IIB string attached to the probe D3-brane, but in the simplifying limit 
the probe D3-brane approaches the $N$ coincident D3-branes. In this case, 
${\bf x}_0 \rightarrow 0$, and the function $G(r)$ in Eq.(2) is reduced to 
\bee
G &=& 1 + g_{\rm eff}^2 \left[ 
\left( {{\sqrt \alpha'} \over r} \right)^4 
+ {k \over N}  
\left( {\sqrt{\alpha'} \over \vert {\bf x}_\perp - {\bf x}_0 \vert } \right)^4
\right]
\nonumber \\
&\rightarrow&
\,\,\, { \widetilde{g_{\rm eff}}^2 \over \alpha'} 
{ 1 \over {\rm U}^4},
\quad {\rm where} \quad
\widetilde{g_{\rm eff}}^2 = \left(1 + {k \over N} \right) g_{\rm eff}^2.
\label{newg}
\eee
The resulting near-horizon geometry is nothing but $AdS_5 \times S^5$ modulo
rescaling of the radius of curvature. Then, the low-energy effective 
Lagrangian Eq.(10) becomes 
\begin{eqnarray}
 L &=& {T_{(n,0)} \over 2} \int d{\rm U} 
\left[ {\rm U}^2 \left( \frac{\widetilde{g_{\rm eff}}^2 }{{\rm U}^4} 
\left(\partial_t {\bf \Omega} \right)^2 - \left(\partial_{\rm U} 
{\bf \Omega} \right)^2 \right)
       + \left(\partial_t {\bf X}_{\parallel} \right)^2 - 
\frac{{\rm U}^4}{\widetilde{g_{\rm eff}}^2} 
\left(\partial_{\rm U} {\bf X}_{\parallel} \right)^2 \right].
 \label{SUGRA}
\end{eqnarray}
Introducing tortoise coordinate $\sigma$ as
\begin{eqnarray}
 \frac{\partial {\rm U}}{\partial \sigma} = \frac{{\rm U}^2}{\widetilde{
g_{\rm eff} }} \qquad \longrightarrow \qquad \frac{1}{\rm U} = 
\frac{\sigma}{\sqrt{\widetilde{ g_{\rm eff}} }} ,
\end{eqnarray} 
and also a dimensionless field variable ${\bf Y}_{\parallel}(t, \sigma)$
\begin{eqnarray}
 {\bf X}_{\parallel} (t, \sigma) = \frac{\sigma}{\widetilde{g_{\rm eff}} } 
{\bf Y}_{\parallel} (t, \sigma),
\end{eqnarray} 
one obtains
\begin{eqnarray}
 L = {T_{(n, 0)} \over 2} \int d \sigma 
\left[ \widetilde{g_{\rm eff}} \left( \left(\partial_t {\bf \Omega} \right)^2 
- \left(\partial_{\sigma} {\bf \Omega} \right)^2 \right) 
+ \frac{1}{\widetilde{g_{\rm eff}}} \left( \left(\partial_t 
{\bf Y}_{\parallel} \right)^2 
- \left( \partial_ {\sigma} {\bf Y}_{\parallel} \right)^2 
- \frac{2}{\sigma^2} {\bf Y}_{\parallel}^2 \right) \right].
\end{eqnarray}
For monochromatic fluctuations ${\bf \Omega} (\sigma, t) 
= {\bf \Omega}(\sigma) e^{- i \omega t}$, ${\bf Y}_{\parallel} 
(\sigma, t) = {\bf Y}_{\parallel} (\sigma) e^{- i \omega t}$,  
the field equations are reduced to one-dimensional Schr\"odinger equations
\begin{eqnarray}
 -\frac{\partial^2}{\partial \sigma^2}  {\bf \Omega} \, \, = 
\,\, \omega^2 {\bf \Omega} \\
 \left(-\frac{\partial^2}{\partial \sigma^2} + {2 \over \sigma^2} \right)  
 {\bf Y}_{\parallel} = \omega^2 {\bf Y}_{\parallel}.
\end{eqnarray}
One thus finds that the macroscopic Type IIB string hovers around on $S^5$ 
essentially via random walk but, on $AdS_5$, fluctuations are mostly 
concentrated on the region $\alpha' {\rm U}^2 \ll \widetilde{g_{\rm eff}}$, 
viz. interior of $AdS_5$.  
 
%%%%%%%%%%%%%%%%%%%%%%%%%%%%%%%%%%%%%%%%%%%%%%%%%%%%%%%%%%%%%%%%%%%%%%%%%%%%%
\section{Strings on D3-Brane: Born-Infeld Analysis}
%%%%%%%%%%%%%%%%%%%%%%%%%%%%%%%%%%%%%%%%%%%%%%%%%%%%%%%%%%%%%%%%%%%%%%%%%%%%%
Let us now turn to world-volume description of semi-infinite strings
ending on D3-branes. From Polchinski's conformal field theory point of view,
which is exact at $\lambda_{\rm IIB} = 0$, the end of fundamental string
represents an electric charge (likewise, the end of D-string represents a
magnetic charge). For semi-infinite string, the electrically charged object
has infinite inertia mass, hence, is identified with a heavy quark $Q$
(or anti-quark $\bar Q$).
An important observation has been advanced recently by Callan and
Maldacena~\cite{callanmaldacena} (and independently by Gibbons~\cite{gibbons}
 and by Howe, Lambert and West~\cite{howe})
that the semi-infinite fundamental string can be realized as a deformation
of the D3-brane world-volume. It was also emphasized by Callan and Maldacena
that full-fledged Born-Infeld analysis is necessary in order to match the 
string dynamics correctly.

In this Section, we reanalyze configuration and low-energy dynamics of
the semi-infinite strings from the viewpoint of deformed world-volume of
D3-branes.
With our ultimate interest to $g_{\rm eff} \rightarrow \infty$ and
zooming into the anti-de Sitter spacetime, we will proceed our analysis
with two different
types of Born-Infeld theory. The first is defined by
the standard Born-Infeld action, which resums (a subset of)
infinite order $\alpha'$ corrections.
Since string loop corrections are completely suppressed, results deduced
from this are only applicable far away from the D3-branes. As such, we will
refer this regime as being described by {\sl classical} Born-Infeld theory.
The second is the conformally invariant Born-Infeld action~\cite{maldacena},
which resums planar diagrams
of `t Hooft's large $N$ expansion in the limit $g_{\rm eff}
\rightarrow \infty$. With the near-horizon geometry fully taken into account,
results obtained from this are directly relevant to the anti-de Sitter
spacetime. We will refer this case as being described by {\sl quantum} 
Born-Infeld theory.

%%%%%%%%%%%%%%%%%%%%%%%%%%%%%%%%%%%%%%%%%%%%%%%%%%%%%%%%%%%%%%%%%%%%%%%%%%%%%
\subsection{Heavy Quark in Classical Born-Infeld Theory}
%%%%%%%%%%%%%%%%%%%%%%%%%%%%%%%%%%%%%%%%%%%%%%%%%%%%%%%%%%%%%%%%%%%%%%%%%%%%%
Classical Born-Infeld theory for D3-branes in flat spacetime is described
by :
\be
L_{\rm CBI} = {1 \over \lambda_{\rm IIB}} \int d^3 x \,
\sqrt{ \det ( \eta_{ab} + \partial_a X_\perp \cdot
\partial_b X_\perp + \alpha' F_{ab}) }.
\nonumber
\ee
For a static configuration whose excitation involes only electric and
transverse coordinate fields, the Lagrangian is reduced to
\be
L_{\rm CBI} \rightarrow {1 \over \lambda_{\rm IIB}}
\int d^3 x \, \sqrt{ (1 - {\bf E}^2) ( 1 + (\nabla X_\perp)^2 ) +
({\bf E} \cdot \nabla X_\perp)^2  - \dot X_\perp^2}.
\label{bilagrangian}
\ee
While the equations of motion for ${\bf E}$ and ${\bf X}_\perp$ derived from
Eq.(\ref{bilagrangian}) are complicated coupled nonlinear equations, for a BPS 
configuration, the nonlinearity simplifies dramatically and reduce to a set
of self-dual equations: 
\be
\nabla X_\perp \cdot {\hat {\bf \Omega}_5} = \pm {\bf E}
\label{cbpscond}
\ee
Here, $\hat {\bf \Omega}_5$ denotes the angular
orientation of the semi-infinite string. The two choices of signs in 
Eq.(\ref{cbpscond}) corresponds to quark and anti-quark and are oriented at anti-podal
points on $\bf \Omega_5$. Once the above BPS condition Eq.(\ref{cbpscond}) is
satisfied, the canonical momentum conjugate to
gauge field reduces to the electric field ${\bf E}$, much as in Maxwell
theory. Moreover, such a solution is a BPS configuration. This follows from
inserting the relation $\nabla X_i = \pm {\bf E}$ into
the supersymmetry transformation of the gaugino field (in ten-dimensional
notation):
\bee
\delta \chi &=& \Gamma^{MN} F_{MN} \epsilon, \qquad \quad 
(M,N = 0, 1, \cdots, 9)
\nonumber \\
&=&  {\bf E} \cdot \Gamma^r \left( \Gamma^0 + {\hat {\bf \Omega}}_5 \cdot \Gamma
\right) \epsilon.
\label{gauginosusy}
\eee
By applying Gauss' law, a semi-infinite strings representing a spherically
symmetric heavy quark or an anti-quark of total charge $n$ is easily 
found~\footnote{
If all the semi-infinte
strings emanate from one of the D3-branes, the center-of-mass
factor $N$ should be absent in the expression.} :
\be
X_\perp \cdot \hat{\bf \Omega}_5 = X_{\perp 0} + \lambda_{\rm IIB} 
{ n \over r} \qquad \left(r = \vert {\bf x}_{\parallel} \vert \right) .
\ee
We emphasize again that 
the BPS condition is satisfied if all the strings (representing
heavy quarks) have the same value of $\Omega_5$ and all the anti-strings
(representing heavy anti-quarks) have the anti-podally opposite value of
$\Omega_5$.

Now that the heavy quarks and anti-quarks are realized as infinite strings,
they can support gapless low-energy excitations. From the D3-brane point
of view, these excitations are interpreted as internal excitations on
${\bf R}_+ \times S_5$. We would like to analyze these low-energy excitations
by expanding the classical Born-Infeld action around a single string
configuration. The expansion is tedious but straightforward.
Fluctuations to quadratic order come from two sources. The first is from
second-order variation of the transverse coordinates.
The second is from square of the first-order variation involving both
transverse coordinates and gauge fields. Evidently, if the background
involves nontrivial transverse coordinate fields, this contribution induces
mixing between gauge field and transverse coordinate fluctuations.
Denoting gauge field fluctuation as ${\cal F}_{\mu \nu}$ and scalar field
fluctuation parallel and perpendicular to the string direction as 
$Y_{\parallel}, Y_\perp$, respectively, the low-energy effective Lagrangian 
is reduced to
\bee
L_{CBI}
= {1 \over 2 \lambda_{\rm IIB} } \int d^3 x \,  \Big[
(1 &+& {\bf E}^2) {\cal F}_{0i}^2  - {\cal F}_{ij}^2
- 2 {\bf E}^2 \, {\cal F}_{0i} \cdot \partial_i Y_{\parallel}
+ {\dot Y_{\parallel}}^2 - (1 - {\bf E}^2) (\partial_i Y_{\parallel})^2
\nonumber \\
&+& (1 + {\bf E}^2) {\dot Y_\perp}^2 - (\partial_i Y_\perp)^2 \, \Big].
\eee
In order to compare the result with supergravity analysis, it is necessary
to integrate out the world-volume gauge fields. The longitudinal scalar
field fluctuation couples only to the electric field. Since the gauge field
fluctuations appear through field strengths, integrating out the gauge
field is straightforward. For the S-wave modes, the reduced Lagrangian reads:
\be
L_{CBI}
= {1 \over 2 \lambda_{\rm IIB}}
\int d^3 x \, \left( (\partial_t Y_{\parallel})^2 - {1 \over (1 + {\bf E}^2)}
(\partial_r Y_{\parallel})^2
+ (1 + {\bf E}^2) (\partial_t Y_\perp)^2 - (\partial_r Y_\perp)^2  \right).
\label{effective}
\ee
The structure of this Lagrangian is quite reminiscent of supergravity
fluctuation Lagrangian Eq.(10) even though
the coordinates involved are quite different. To make further comparison,
we first note that the world-volume coordinate $x$ is {\sl not} the intrinsic
coordinates measured {\sl along} the D3-brane world-volume. Since we are
studying fluctuation on the D-brane, it is quite important to measure distance
using intrinsic D3-brane coordinates.
Therefore, we now make a change of variable $r$ to the tortoise
coordinate $\sigma$ :
\be
{d r \over d \widetilde \sigma} = {1 \over \sqrt {\widetilde G}} ;
\hskip1.5cm {\tilde G}(r) \equiv (1 + {\bf E}^2) = \left( 1 +
{ n^2 \lambda_{\rm IIB}^2 \over r^4} \right)
\ee
After the change of variables, Eq.(\ref{effective}) becomes:
\be
 L_{\rm CBI} = {1 \over 2 \lambda_{\rm IIB}}
\int d {\tilde \sigma} \, r^2 \,
\left[ {\sqrt {\widetilde G}} \Big( (\partial_t Y_\perp)^2
- (\partial_{\tilde \sigma} Y_\perp)^2 \Big)
+ {1 \over {\sqrt {\widetilde G}}} \Big( (\partial_t Y_{\parallel})^2
- (\partial_{\tilde \sigma} Y_{\parallel})^2
\Big) \, \right].
\ee
Again, the Lagrangian clearly displays the fact that D3-brane coordinate
fluctuations parallel and perpendicular to the semi-infinite string
propagates at the speed of light even though string mass density and
tension changes spatially. Moreover, polarization dependence of string mass
density and tension can be understood geometrically from the
fact that the proper parallel and orthogonal directions to the D-brane 
does not coincide with the above fixed background decomposition.
In fact, this has been demonstrated explicitly for the case of open string 
ending on D1-brane case~\cite{reyyee}. Since essentially the same analysis 
is applicable for D3-brane, we will not elaborate on
it further here and move on to the analysis of boundary conditions.

For a monochromatic transverse fluctuation $Y_\perp (\tilde \sigma, t)
= Y_\perp (\tilde \sigma) e^{- i \omega t}$, unitary transformation
$Y_\perp \rightarrow Y_\perp/ r G^{1/4}$ and change of variables
$\tilde \sigma \rightarrow \tilde \sigma / \omega, \, 
r \rightarrow r / \omega, \, \lambda_{\rm IIB} \rightarrow 
\lambda_{\rm IIB} / \omega$ yields the fluctuation
equation of motion into the form of a one-dimensional Schr\"odinger equation:
\be
\left[ - {d^2 \over d \tilde \sigma^2} + \tilde{ V_\perp} (\tilde \sigma) 
\right] Y_\perp (\tilde \sigma)
= + 1 \cdot Y_\perp ( \tilde \sigma),
\ee
where
\be
\hskip1cm \tilde {V_\perp} (\tilde \sigma) = {5 {\tilde \epsilon}^{-2} 
\over (\tilde \sigma^2 / \tilde \epsilon^2 +
\tilde \epsilon^2 /\tilde \sigma^2)^3} ; 
\hskip1.5cm \left(\tilde \epsilon = \sqrt{n \lambda_{\rm IIB}} \omega
\right).
\ee
Note that the functional form of this equation is exactly the same as
one obtained from supergravity description.
Therefore, the fact that the self-adjoint boundary condition of the
$Y_\perp$ fluctuation is Dirichlet type holds the same.

Repeating the analysis for monochromatic parallel fluctuations
$Y_{\parallel} (\tilde \sigma, t) = Y_{\parallel} (\tilde \sigma)
e^{- i \omega t}$, unitary transformation $Y_{\parallel} \rightarrow
r^{-1} G^{1/4} Y_{\parallel}$ and the same change of variables as above yields
analog one-dimensional Schr\"odinger equation:
\be
\left[ -{d^2 \over d \tilde \sigma^2} + \tilde{ V_{\parallel}} (\tilde
\sigma) \right] \, Y_{\parallel} (\tilde \sigma) 
= + 1 \cdot Y_{\parallel} (\tilde \sigma)
\ee
where 
\be
\tilde{ V_{\parallel}} (\tilde \sigma)  = {(6 \tilde \epsilon^2 / \tilde r^2
- \tilde r^2 / \tilde \epsilon^2)
\over
(\tilde r^2 / \tilde \epsilon^2 + \tilde \epsilon^2 / \tilde r^2)^3 }.
\ee
Comparison to result Eq.(17) shows that, once again, the functional behavior
is essentially the same between the supergravity and the classical Born-Infeld 
side. 
As such, for low-energy and weak string coupling $g_{\rm IIB} \rightarrow 0$,
both sides gives rise now to Neumann boundary condition, which is another 
possible self-adjoint extension of one-dimensional wave operator.
Quite surprisingly, we have reproduced the Polchinski's boundary condition 
for an open string ending on D3-branes purely from dynamical considerations 
both in spacetime (using supergravity description) and on D3-brane worldvolume
(using Born-Infeld description).

%%%%%%%%%%%%%%%%%%%%%%%%%%%%%%%%%%%%%%%%%%%%%%%%%%%%%%%%%%%%%%%%%%%%%%%%%%%%
\subsection{Heavy Quark in Quantum Born-Infeld Theory}
%%%%%%%%%%%%%%%%%%%%%%%%%%%%%%%%%%%%%%%%%%%%%%%%%%%%%%%%%%%%%%%%%%%%%%%%%%%%
In the regime $g_{\rm eff} \rightarrow \infty$, the D3-brane dynamics is
most accurately described by quantum Born-Infeld theory, in which `t Hooft's
planar diagrams are resummed over. One immediate question is whether and how
the shape and fluctuation dynamics of semi-infinite string are affected by
these quantum corrections. To answer this question, we analyze semi-infinite
string configuration ending on a D3-brane located in the vicinity of other
$N-1$ D3-branes. The configuration is depicted in Fig. 1.

\begin{figure}[t]
   \vspace{0cm}
   \epsfysize=7cm
   \epsfxsize=13cm
   \centerline{\epsffile{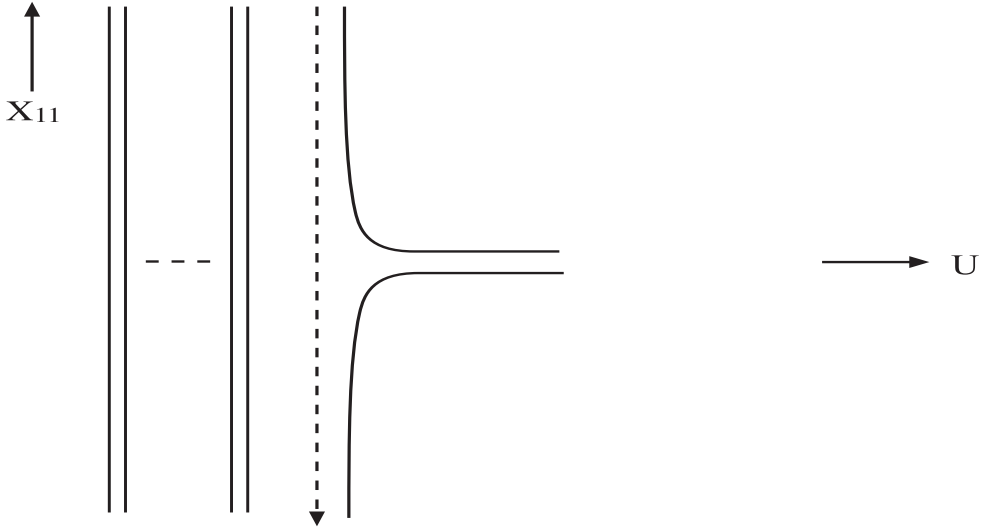}}
\caption{Macroscopic string as a BPS soliton on D3-brane worldvolume.
Large-N corrections induced by branes at $U = 0$ in general gives rise
to corrections to the shape and low-energy dynamics of the D3-brane.}
\end{figure}

The quantum Born-Infeld theory is described by the Lagrangian
\bee
L_{\rm QBI} &=& {1 \over \lambda_{\rm IIB}}
\int d^3 x {1 \over h} \left[ \sqrt{ \det \left(
\eta_{ab} + h (\partial_a {\bf X}_\perp \cdot \partial_b {\bf X}_\perp)
+ \sqrt{h} \, F_{ab} \right) }  - 1 \right]
\nonumber \\
h ({\rm U}) &=& {g_{\rm eff}^2 \over {\rm U}^4} \, ; 
\hskip2cm \left({\rm U} = |{\bf X}_\perp| / \ell_s^2 \right) \, .
\nonumber 
\eee
The $-1$ term inside the bracket originates from the Wess-Zumino term of 
D-brane worldvolume action and ensures that the ground state has zero energy.
For a static worldvolume configuration with nontrivial electric and U-fields,
one finds
\bea
L_{\rm QBI} = {1 \over \lambda_{\rm IIB}}
 \int d^3 x {1 \over h} \left[ \sqrt{
\left( 1 - h {\bf E}^2 \right)
\left( 1 + h (\nabla {\rm U})^2 \right) +
h^2 ({\bf E} \cdot \nabla {\rm U})^2 -
h  \dot {\rm U}^2 } - 1 \right].
\label{qbi}
\eea
Denoting the quantity inside the square root as $L$ for notational
brevity, the canonical conjugate momenta to the gauge field and the
Higgs field U are given by:
\bee
\lambda_{\rm IIB} {\bf \Pi}_A &=&
{1 \over L }
\left[ - {\bf E} \left( 1 + h (\nabla {\rm U})^2 \right)
+ h \nabla U ( {\bf E} \cdot \nabla U) \right]
\nonumber \\
\lambda_{\rm IIB} {\bf P}_{\rm U}  &=& - {1 \over L } \dot{\rm U}.
\eee
We now look for a BPS configuration of worldvolume deformation, 
as in the case of the classical Born-Infeld theory, that can be interpreted 
as a semi-infinte string attached to the D3-branes. For a static 
configuration, the equations of motions read:
\bee
&& \nabla \cdot \left[{1 \over L } 
\Big( \nabla {\rm U} ( 1 - h {\bf E}^2 )
+ h ({\bf E} \cdot \nabla {\rm U}) {\bf E} \Big) \right]
= {4 {\rm U}^3 \over L} h
\Big[ ({\bf E} \cdot \nabla {\rm U})^2 - {\bf E}^2 (\nabla {\rm U})^2 \Big],
\nonumber \\
&& \nabla \cdot \Big[ {1 \over h L }
\Big( - {\bf E} ( 1 + h (\nabla {\rm U})^2 )
+ h \nabla {\rm U} ({\bf E} \cdot \nabla {\rm U}) \Big) \, \Big] = 0.
\label{qbieom}
\eee
While coupled in a complicated manner, it is remarkable that the two 
equations can be solved exactly by the following self-dual BPS equation:
\be
{\bf E} = \pm \nabla {\rm U}.
\label{bps}
\ee
Remarkably, this self-dual equation is exactly of the 
same form as the one found for the classical Born-Infeld theory, Eq.(20). 
In this case, $L = 1/h$ and
nonlinear terms in each equations cancel each other. 
We emphasize that the Wess-Zumino term $-1$ in the quantum Born-Infeld 
Lagrangian, which were present to ensure vanishing ground-state energy, is 
absolutely crucial to yield the right-hand side of the first equation of 
motion, Eq.(\ref{qbieom}). 
The resulting equation is nothing but Gauss' law constraint, Eq.(\ref{qbieom}):
\be
\nabla \cdot {\bf E} =  \nabla^2 {\rm U} = 0,
\label{qbisd}
\ee
where the Laplacian is in terms of conformally flat coordinates. 
Spherically symmetric solution of the Higgs field U is given by 
\be
{\rm U} = {\rm U}_0 + \lambda_{\rm IIB} {n \over 
r}, \hskip1.5cm (r = |{\bf x}_{\parallel} |).
\label{higgsfield} 
\ee
Interpretation of the solution is exactly the same as in the classical 
Born-Infeld theory:
gradient of the Higgs field U acts as a source of the 
world-volume electric field. 
See Eq.(\ref{qbisd}). From the Type IIB string theory point of view, the 
source is nothing but $n$ coincident Type IIB fundamental strings attached 
to the D3-branes. As such, one now has found a consistent worldvolume 
description of the macroscopic Type IIB string in the `t Hooft limit. 

The total energy now reads
\bee
E &=& \int d^3 x \, \left( {1 \over h} [ 1 + h (\nabla {\rm U})^2]
- {1 \over h} \right)
\nonumber \\
&=& \int d^3 x \, (\nabla {\rm U})^2
\nonumber \\
&=& n \, {\rm U}(r=\epsilon).
\eee
Thus, the total energy diverges with the short-distance cut-off $\epsilon$
as in the weak coupling case. 
Since the above spike soliton is a BPS state and has 
a nonsingular tension the solution remains valid even at 
strong coupling regime. 

%%%%%%%%%%%%%%%%%%%%%%%%%%%%%%%%%%%%%%%%%%%%%%%%%%%%%%%%%%%%%%%%%%%
\subsection{Quantum Born-Infeld Boundary Condition}
%%%%%%%%%%%%%%%%%%%%%%%%%%%%%%%%%%%%%%%%%%%%%%%%%%%%%%%%%%%%%%%%%%%
We will now examine fluctaution of the Born-Infeld fields in the quantum
soliton background. The setup is as in the previous subsection --- the $N$ 
multiple $D3$-branes produces the $AdS_5$ backgound, and worldvolume dynamics
of a single $D3$-brane in this background is described by the quantum 
Born-Infeld theory, Eq.(\ref{qbi}). Keeping up to harmonic terms, 
the fluctuation Lagragian becomes 
\begin{eqnarray}
 L^{(2)} &=& -\frac{1}{\lambda_{\rm IIB}} \int d^3 r \frac{1}{2} 
	\left[ F_{\alpha \beta}^2 - (1+ \frac{g_{\rm eff}^2}{{\rm U}^4}
	(\partial_r {\rm U})^2 ) F_{0 \alpha}^2  
   	- (\partial_0 \chi)^2 + \left(1-\frac{g_{\rm eff}^2}{{\rm U}^4}
        (\partial_r {\rm U})^2 \right) (\partial_\alpha \chi)^2 \right.
\nonumber \\ 
   & &	\left. + 2 \frac{g_{\rm eff}^2}{{\rm U}^4}
        (\partial {\rm U})^2 F_{0 \alpha} \partial_\alpha \chi  
   	+ 12 \frac{{\rm U}^2}{g_{\rm eff}^2} \chi^2  
    	+ {\rm U}^2 \left( - (1 +\frac{g_{\rm eff}^2}{{\rm U}^4} 
          (\partial {\rm U})^2 \right) (\partial_0 \theta)^2 
	+ (\partial_\alpha \theta)^2 ) \right], 
\end{eqnarray}
where $\chi$ refers to the radial direction fluctuation, and $\psi$ is 
the angular fluctuation corresponding to the coordinates $\theta$ in the 
lagrangian Eq.(\ref{qbi}). With the Higgs field given as in 
Eq.(\ref{higgsfield}), the above fluctuation Lagragian is complicated. 
Thus, we will consider a special situation, for which ${\rm U}_0 = 0$. 
In this case, one finds that
\begin{eqnarray}
 \frac{g_{\rm eff}^2}{{\rm U}^4}(\partial_r {\rm U})^2 = 
\frac{g_{\rm eff}^2}{\lambda_{\rm IIB}^2 n^2}.
\end{eqnarray}
This simplifies the fluctuation Lagrangian considerably, yielding
\begin{eqnarray}
 L^{(2)} &=& -\frac{1}{\lambda_{\rm IIB} } \int d^3 r \frac{1}{2}
 \left[ F_{\alpha \beta}^2 - \left(1+ \frac{g_{\rm eff}^2}{q^2} \right) F_{0 \alpha}^2  
-(\partial_0 \chi)^2 + \left(1-\frac{Q^2}{q^2} \right)(\partial_\alpha \chi)^2 
\right.
\nonumber \\
   & & \left. + 2 \frac{g_{\rm eff}^2}{\lambda_{|rm IIB}^2 n^2} 
          F_{0 \alpha} \partial_\alpha \chi
        + 12 \frac{{\rm U}^2}{g_{\rm eff}^2} \chi^2  
        + {\rm U}^2  ( - (1 + \frac{g_{\rm eff}^2}{\lambda_{\rm IIB}^2 n^2}) 
          (\partial_0 \theta)^2
        + (\partial_\alpha \theta)^2 ) \right].
\nonumber
\end{eqnarray}
One readily finds that the electric field and the radial Higgs field fluctuations are related each other by
\begin{eqnarray}
 (1+ \frac{g_{\rm eff}^2}{\lambda_{\rm IIB} n^2}) F_{0 \alpha} 
= \frac{g_{\rm eff}^2}{\lambda_{\rm IIB} n^2} 
\partial_\alpha \chi.
\nonumber
\end{eqnarray}
As such, integrating out the electric field fluctuation, we find that
\begin{eqnarray}
 L^{(2)} &=& - \frac{1}{\lambda_{\rm IIB}} \Omega_2 \int dr r^2 
\frac{1}{2} \left[ F_{\alpha \beta}^2
	 -(\partial_0 \chi)^2 + 
	\frac{1}{1+g_{\rm eff}^2/\lambda_{\rm IIB} n^2} 
        (\partial_\alpha \chi)^2 + 12 \frac{\lambda_{\rm IIB} n^2}
        {g_{\rm eff}^2} \frac{1}{r^2} \chi^2 \right. \nonumber \\
	&& - \left. \frac{\lambda_{\rm IIB}^2 n^2}{r^2}
        \left(- \left(1+ {g_{\rm eff}^2 \over \lambda^2_{\rm IIB} n^2} \right) 
        (\partial_0 \theta)^2 + (\partial_\alpha \theta)^2 \right) \right],
\qquad \qquad (\Omega_2 \equiv {\rm Vol}(S_2)).
\nonumber
\end{eqnarray}
We see that fluctuation of the magnetic field is non-interacting, and hence
focus on the Higgs field fluctuations only. Make the following change of 
radial coordinate and Higgs field \footnote{The change of variable for 
$\chi$ field renders $\tilde \chi$ dimensionless.}:
\begin{eqnarray}
 r = \frac{1}{\sqrt{1 + {g_{\rm eff}^2 \over \lambda_{\rm IIB}^2 n^2}} } 
\tilde{r} \qquad {\rm and} \qquad \chi =  \lambda_{\rm IIB} \frac{n}{r} 
\tilde{\chi} = {\rm U} \widetilde{\chi}. 
\end{eqnarray}
The fluctuation Lagrangian then becomes
\begin{eqnarray}
 L^{(2)} &=&  \frac{1}{\lambda_{\rm IIB}} \Omega_2 
         \int d\tilde{r} \, \frac{1}{2} q^2
          \frac{1}{\sqrt{1 + g_{\rm eff}^2 /\lambda_{\rm IIB} n^2}} 
         \left[ (\partial_0 \tilde{\chi})^2 - 
        (\partial_{\tilde{r}}  \tilde{\chi})^2 -12 
        \left( \frac{\lambda_{\rm IIB}^2 n^2}{g_{\rm eff}^2} + 1 \right)
        \frac{\tilde{\chi}^2}{\tilde{r}^2} \right] \nonumber \\
        & +&\frac{1}{\lambda_{\rm IIB}} \Omega_2 \int d\tilde{r} \,
         \frac{1}{2} \lambda_{\rm IIB}^2 n^2 
         \sqrt{1 + g_{\rm eff}^2/ \lambda_{\rm IIB}^2 n^2} 
       \left[ (\partial_0 \theta)^2 - (\partial_{\tilde{r}} \theta)^2  \right].
\nonumber
\end{eqnarray}
The overall $\lambda_{\rm IIB}^2 n^2$ factor is actually irrelevant, as 
it can be eliminated by redifining the $\theta$ and $\tilde{\chi}$ fields
appropriately. With an appropriate change of variables as in the
supergravity case, we finally obtain the fluctuation equations of motion as:
\begin{eqnarray}
 \left[ -\frac{\partial^2}{\partial_{\tilde r}^2} - \omega^2 \right] \theta 
&=& 0
\nonumber \\
 \left[-\frac{\partial^2}{\partial_{\tilde r}^2} + 12 \frac{{\rm U}^2}{g_{\rm eff}^2}  - \omega^2 \right] \tilde \chi &=& 0.
\nonumber 
\end{eqnarray}
Remarkably, while not transparent in the intermediate steps, the Higgs field 
fluctuations turn out to be independent of the $\lambda_{\rm IIB} n$ parameter. 
It implies that the flucutations exhibit a universal dynamics, independent 
of magnitude of the `quark' charge. 
The fluctuations comprise essentially of the Goldstone modes on $S_5$ and 
harmonically confined radial Higgs field fluctuation localized near $u=0$. 
Implication of these characteristics of the fluctuations to the super 
Yang-Mills theory is discussed elsewhere \cite{rty}.
%%%%%%%%%%%%%%%%%%%%%%%%%%%%%%%%%%%%%%%%%%%%%%%%%%%%%%%%%%%%%%%%%%%%%%%%%%%%
\subsection{Geometric UV-IR Duality}
%%%%%%%%%%%%%%%%%%%%%%%%%%%%%%%%%%%%%%%%%%%%%%%%%%%%%%%%%%%%%%%%%%%%%%%%%%%%
It is remarkable that, for both the supergravity and the Born-Infeld theory
viewpoints, the fluctuation dynamics is identical
given the fact that $\sigma$ tortoise coordinate in the supergravity
description measures the distance along $\alpha'$U direction --
a direction {\sl perpendicular} to the D3-brane, while $\tilde \sigma$
tortoise coordinate in the classical Born-infeld description measures the
distance {\sl parallel} to the D3-brane -- Yang-Mills distance.
The supergravity and the classical Born-Infeld theory provides dual
description of the semi-infinte string as a heavy quark. The reason behind
this is that, as $\alpha'$
corrections are taken into account, the D3-brane is pulled by the semi-infinite
string and continue deforming until tensional force balance is achieved.
Now that D3-brane sweeps out in $\alpha'$U direction once stretched by
charge probes, balance of tensional force relates
\be
{1 \over R_{\parallel}} \hskip0.5cm \leftrightarrow \hskip0.5cm
\alpha' {\rm U},
\label{uvir}
\ee
where $R_{\parallel} = \vert {\bf x}_{\parallel} \vert$. In particular, the
short (long) distance in directions parallel to the D3-brane is related
to long (short) distance in direction perpendicular to the D3-brane.

We will refer the `reciprocity relation' Eq.(\ref{uvir}) as `geometric UV-IR
duality' and will derive in later sections
a precise functional form of the relation from the
consideration of quark-antiquark static energy.

%%%%%%%%%%%%%%%%%%%%%%%%%%%%%%%%%%%%%%%%%%%%%%%%%%%%%%%%%%%%%%%%%%%%%%%%%%%
\section{String Anti-string Pair and Heavy Quark Potential}
%%%%%%%%%%%%%%%%%%%%%%%%%%%%%%%%%%%%%%%%%%%%%%%%%%%%%%%%%%%%%%%%%%%%%%%%%%%
So far, in the previous sections, 
we have studied BPS dynamics involving a single probe string. 
In this section, we extend the study to non-BPS configuration.
We do this again from Born-Infeld super Yang-Mills and anti-de Sitter
supergravity points of view. 
Among the myriad of non-BPS configurations, the simplest and physically
interesting one is a pair of oppositely oriented, semi-infinte strings
attached to the D3-brane.
 
Physically, the above configuration may be engineered as follows. We first
prepare a macroscopically large, U-shaped fundamental string, whose tip 
part is parallel to the D3-brane but the two semi-infinite sides are
oriented radially outward. See figure 2. 
As we move this string toward D3-brane, the
tip part will be attracted to the D3-brane and try to form a non-threshold
bound-state. The configuration is still not a stable BPS configuration
since the two end points from which semi-infinite sides emanate acts as
a pair of {\sl opposite} charges since their $\Omega_5$ orientation is 
the same. They are nothing but heavy quark anti-quark pairs. As such, the 
two ends will attract each
other (since the bound-state energy on the D3-brane is lowered by 
doing so) and eventually annihilate into radiations. However, in so far
as the string is semi-infinite, the configuration will be energetically
stable: inertia of the two open strings is infinite. Stated differently,
as the string length represents the vacuum expectation value of Higgs field, 
the quark anti-quark pairs are infinitely heavy. In this way, we have
engineered static configuration of a $(Q \overline{ Q})$ pair on the D3-brane. 

\begin{figure}[t]
   \vspace{0cm}
   \epsfysize=7cm
   \epsfxsize=13cm
   \centerline{\epsffile{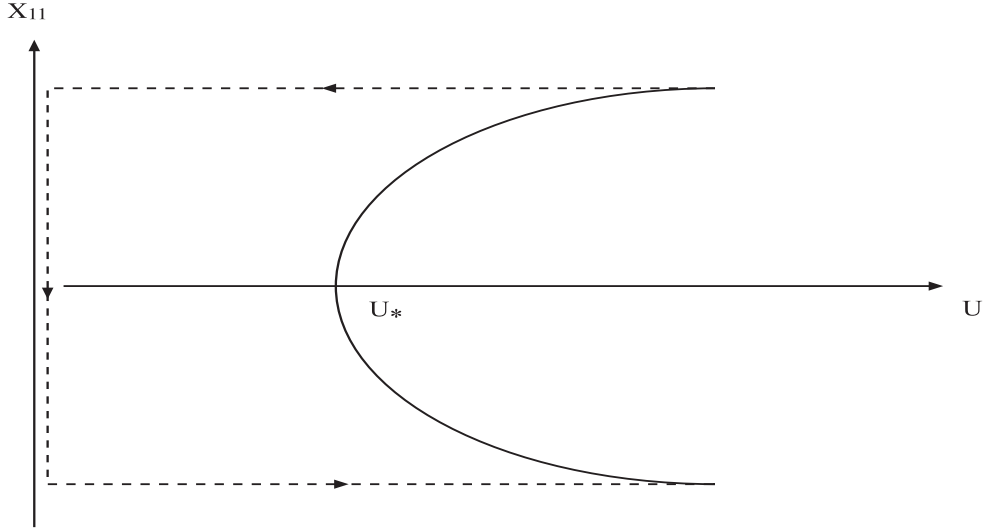}}
\vspace{0.5cm}
\caption{Non-BPS configuration of string anti-string pair as realization of
heavy quark anti-quark pair. String corrections smooth out curvature at the
two sharp corners.}
\end{figure}

The $(Q \overline{ Q})$ configuration is of some interest since it may tell us 
whether the $d=4, {\cal N}=4$ super Yang-Mills theory exhibits confinement. 
The theory
has a vanishing $\beta$-function, hence, no dimensionally transmuted mass 
gap either. As such, one might be skeptical to a generation of a physical scale
from {\sl gedanken} experiment using the above configuration. The result we
will be getting is not in contradiction, however, as the scale interpreted as
a sort of `confinement' scale is really residing in $AdS_5$ spacetime. It is
a direct consequence of spontaneously broken conformal invariance of the 
super Yang-Mills theory. Therefore, the `confinement' behavior in $AdS_5$ spacetime
ought to be viewed as `Coulomb' behavior in super Yang-Mills theory. Once 
again, the interpretation relies on the earlier
observation that paralle and perpendicular directions to D3-brane are
geometrically dual each other.
%%%%%%%%%%%%%%%%%%%%%%%%%%%%%%%%%%%%%%%%%%%%%%%%%%%%%%%%%%%%%%%%%%%%%%%%%%
\subsection{Quark-Antiquark Pair: String in Anti-de Sitter Space}
%%%%%%%%%%%%%%%%%%%%%%%%%%%%%%%%%%%%%%%%%%%%%%%%%%%%%%%%%%%%%%%%%%%%%%%%%%
We first construct the aforementioned string configuration corresponding
to $Q \overline{Q}$ pair on D3-brane from anti-de Sitter supergravity. To 
find the configuration we find it most convenient to study portions of
the string separately.
Each of the two semi-infinite portions is exactly the same as a single 
semi-infinite string studied in the previous section. Thus, we concentrate
mainly on the tip portion that is about to bound to the D3-brane. The
portion cannot be bound entirely parallel to the D3-brane since it will 
cause large bending energy near the location we may associate with $Q$ and
$\overline{Q}$. The minimum energy configuration would be literally like U-shape.
We now show that this is indeed what comes out.

We now repeat the analysis of test string in supergravity background of
$N$ D3-branes. For a static configuration, the Nambu-Goto Lagrangian is
exactly the same as Eq.(7):
\be
L_{\rm NG} \rightarrow \int d \sigma \, \sqrt { {{\bf X}'_\perp}^2
+ {1 \over G} {{\bf X}'_{\parallel}}^2 }.
\ee
From the equations of motion, we find that the other possible solution is
when the string is oriented parallel to the D3-brane. This yields precisely the
static gauge configuration
\be
X^0 = t = \tau \, ,\hskip0.75cm {\bf X}_{\parallel} = \sigma {\hat {\bf n}}.
\ee
Then, the two equations of motion Eq.(8) become
\bee
\left( {\bf X}'_\perp \over \sqrt{ {{\bf X}'_\perp}^2 + G^{-1} }
\right)' &=& (\nabla_{ {\bf x}_\perp} G^{-1} ) 
\nonumber \\
\left( G^{-1} \over \sqrt{ {{\bf X}'_\perp}^2 + G^{-1} }
\right)' &=& 0.
\label{paralleleq}
\eee
We now consider the non-BPS $Q \overline{Q}$ configuration studied earlier. Since
the two semi-infinite strings are oriented parallel on $\Omega_5$ we only
consider excitation of $\alpha'$U coordinate.
From the equation of motion, the first of Eq.(\ref{paralleleq}),
\be
- {1 \over G} {\rm U}'' + {1 \over 2} \left( \partial_{\rm U} 
{1 \over G} \right) \left( 2 {{\rm U}'}^2 + {1 \over G} \right) = 0,
\ee
one can obtain the first integral of motion:
\be
G^2 {{\rm U}'}^2 + G = { g_{\rm eff}^2 \over {\rm U}^4_*},
\label{tobesolved}
\ee
where we have chosen convenient parametrization of an integration constant.
This is in fact the same as the other conserved integral, the second of
Eq.(\ref{paralleleq}) and shows that the equations are self-consistent.

Denoting $Z \equiv {\rm U}_* / {\rm U}$, 
the solution to Eq.(\ref{tobesolved}) can be
found in an implicit functional form:
\be
(x_{\parallel} - d/2) = \pm { g_{\rm eff} \over {\rm U}_*} \left[
\,\, \sqrt{2} E\left(\arccos Z, {1 \over \sqrt 2} \right) - {1 \over \sqrt 2}
F \left(\arccos Z, 1/{\sqrt 2} \right) \,\, \right].
\label{solution}
\ee
Here, $F(\phi, k), E(\phi, k) $ denote
the elliptic integrals of first and second kinds.
It is easy to visualize that the solution describes
monotonic lifting of $U$-direction fluctuation (thus away from the
D3-brane plane) and diverges at finite distance along $x_{\parallel}$.
For our choice, they are at $x_{\parallel} = 0$ and $d$. This
prompts to interpret the integration constant $d$ in Eq.(\ref{solution}) as
the separation between quark and anti-quark, measured in $x_{\parallel}$
coordinates.
The string is bended (roughly in U-shape) symmetrically about
$x_{\parallel} = d/2$. As such, the inter-quark distance measured
{\sl along} the string is not exactly the same as $d$. The proper distance
along the string is measured by the $U$-coordinate. The relation 
between the coordinate separation and proper separation is obtained
easily by integrating over the above Eq.(48). It yields 
\bee
{d \over 2} &=& {g_{\rm eff} \over {\rm  U}_*} \left[
\sqrt{2} E \left({\pi \over 2}, {1 \over \sqrt 2} \right)
- {1 \over \sqrt 2} F \left( {\pi \over 2}, {1 \over \sqrt 2} \right)
\right]
\nonumber \\
&=& {g_{\rm eff} \over {\rm U}_*} C_1 \, , \hskip2cm \left(C_1 = 
{\sqrt \pi} \Gamma(3/4)/\Gamma(1/4) = 0.59907.... \right) 
.
\eee
This formula implies that the integration constant ${\rm U}_*$ would be
interpreted as the height of the U-shaped tip along U-coordinate. 
Up to numerical factors, the relation again exhibits the `geometric duality'
Eq.(31) between Yang-Mills coordinate distance $d$ and proper distance 
${\rm U}_*$. 

Using the first integral of motion, the inter-quark potential is
obtained straightforwardly from the Born-Infeld Lagrangian. The
proper length of the string is infinite, so we would expect linearly
divergent (in $\rm U$-coordinate) energy. Thus, we first calculate 
regularized expression of energy by excising out a small neighborhood
around $x_{\parallel} = 0, d$:
\bee
V_{Q \overline Q} (d)  &=& {\rm lim}_{\epsilon \rightarrow 0} \, n \, \Big[
{\sqrt {G_*}} \int_0^{d/2 - \epsilon} d x_{\parallel} \, G^{-1} \Big]
\nonumber \\ 
&=& {\rm lim}_{{\rm U} \rightarrow \infty}   \, n \,
\Big[ 2 {\rm U}_* \int_1^{\rm U} dt { t^2 \over \sqrt{t^4 - 1}} \Big]
\nonumber \\
&=& 2 n {\rm U}_* \Big[ {\rm U} + {1 \over \sqrt 2} K( 1/\sqrt 2) - {\sqrt 2}
E(1/\sqrt 2) + {\cal O}(U^{-3}) \Big].
\eee
The last expression clearly exhibits the infinite energy being originated
from the semi-infinite strings and indeed is proportional to the proper 
length 2U. After subtracting (or renormalizing) the string self-energy, 
the remaining, finite part may now be interpreted as the inter-quark 
potential. Amusing fact is that it is proportional to the inter-quark 
distance when measured in $U$-coordinate. One might be tempted to interpret
that the inter-quark potential is in fact a Coulomb potential by using the 
relation Eq.(50). However, it does {\sl not} have the expected dependence on
the electric charges: instead of quadratic dependence, it only grows 
linearly. Because of this, we suspect that the interpretation of static
$Q \overline{Q}$ potential is more natural when viewed as a linearly confining 
potential in U-direction in $AdS_5$. 

The static inter-quark potential shows several peculiarities. First, the
potential is purely Coulombic, viz. inversely proportional to the separation
distance. This, however, is due to the underlying conformal invariance. 
Indeed, at the critical point of second-order phase transition (where conformal
invariance is present), it was known that the Coulomb potential is the
only possible behavior~\cite{peskin}. Second, most significantly, the 
static quark potential strength is {\sl non-analytic} in the
effective `t Hooft coupling constant, $g_{\rm eff}^2$. The quark potential is 
an experimentally verifiable physical quantity, and, in weak `t Hooft coupling
domain, it is well-known that physical quantities ought to be analytic in
$g_{\rm eff}^2$, at least, within a finite radius of convergence around the
origin. Moreover, for d=4, ${\cal N}=4$ super Yang-Mills theory, we do not
expect a phase transition as the `t Hooft coupling parameter is varied. Taking
then the aforementioned nonanalyticity of square-root brach cut type as a 
prediction to the strongly coupled super Yang-Mills theory, we conjecture 
that there ought to be two distinct strong coupling systems connected smoothly 
to one and the same weakly coupled super Yang-Mills theory. To what extent
these two distinct sytems are encoded into a single $AdS_5$ supergravity is
unclear, and hence poses an outstanding issue to be resolved in the future.
%%%%%%%%%%%%%%%%%%%%%%%%%%%%%%%%%%%%%%%%%%%%%%%%%%%%%%%%%%%%%%%%%%%%%%%%%%%
\subsection{Heavy Quark Anti-Quark Pair: Quantum Born-Infeld Analysis}
%%%%%%%%%%%%%%%%%%%%%%%%%%%%%%%%%%%%%%%%%%%%%%%%%%%%%%%%%%%%%%%%%%%%%%%%%%%
Let us begin with quantum Born-Infeld analysis of the heavy quark and 
anti-quark pair.
In earlier Sections, we have elaborated that quarks and anti-quarks correspond
to semi-infinite strings of opposite $\Omega_5$ orientation angle. That this is 
BPS configuration can be understood in several different ways. Consider a 
string piercing the D3-brane radially. The simplest is from the gaugino
supersymmetry transformation, Eq.(\ref{gauginosusy}). 
Residual supersymmetry is consistent 
among individual semi-infinite strings if and only if their $\Omega_5$ angular
orientations are all the same for same charges and anti-podally opposite for
opposite charges.  Alternatively, at the intersection locus, one can 
split the string and slide the two ends in opposite directions. This does not
cost any energy since the attractive electric force is balanced by repulsive
$\alpha'$ U gradient force.  This BPS splitting naturally gives rise to  
a quark anti-quark configuration in which semi-infinite strings are anti-podally
opposite on $\Omega_5$.

The fact that $Q \overline{Q}$ does not exert any force in this case is not a 
contradiction at all. The Coulomb force between $Q$ and $\overline{Q}$ is 
cancelled 
by gradient force of $\alpha'$ U field. This already indicates that we have to
be careful in interpreting the evolution of $Q \overline{Q}$ on the D3-brane
as a timelike Wilson loop of the four-dimensional gauge fields only. The more 
relevant quantity is the full ten-dimensional Wilson loop:
\be
W[C] = \exp \big[ i \oint ( A_\alpha d x^\alpha + 
{\dot {\bf X}}_\perp \cdot d x_\perp ) \big].
\ee
From BPS point of view, it simply states that, for example, in evaluating a 
static potential between heavy quark and anti-quark,
one has to include {\sl all} long-range fields that will produce the potential.

A little thought concerning the BPS condition Eq.(\ref{bps}) 
indicates that there is
yet another configuration that may be interpreted as static $Q \overline{Q}$ 
state.
If we take a semi-infinite string representing a quark with the positive
sign choice in Eq.(36) and superimpose to another semi-infinite string
representing an anti-quark with the negative sign choice, then we obtain 
$Q \overline{Q}$ 
configuration in which the $\Omega_5$ angular positions are the {\sl same}.
In this case, it is easy to convince oneself that both the Coulomb force
and the U-field gradient force are attractive, hence, produce a
nontrivial $Q \overline{Q}$ static potential.
Indeed, starting from the BPS $Q \overline{Q}$ configuration with opposite
$\Omega_5$ orientations mentioned just above, 
one can deform into the present non-BPS $Q \overline{Q}$ configuration
by rotating one of the semi-infinite string on $\Omega_5$ relative to 
the other. See figure 3 for illustration. 
It should be also clear that it is the gradient force of scalar fields on
transverse directions that changes continuously as the relative $\Omega_5$
angle is varied.

\begin{figure}[t]
   \vspace{0cm}
   \epsfysize=7cm
   \epsfxsize=13cm
   \centerline{\epsffile{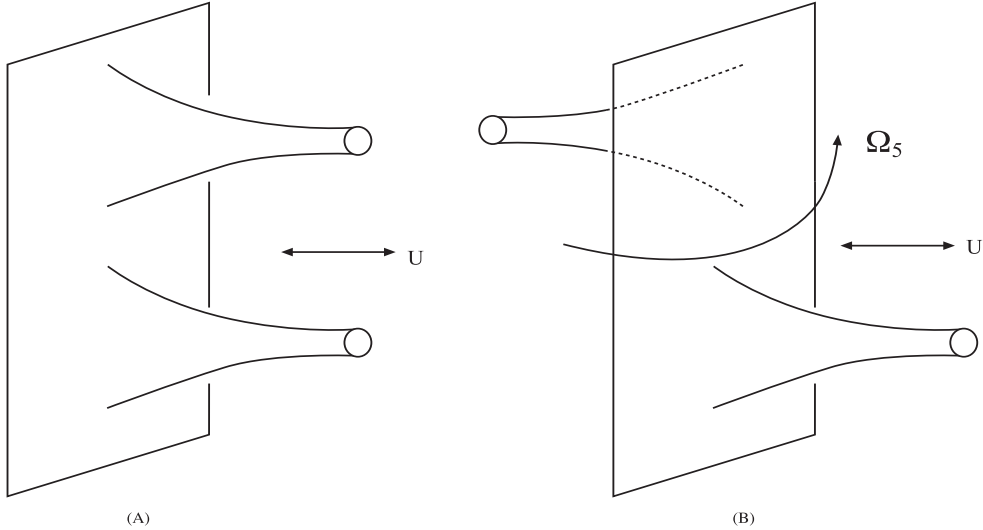}}
\vspace{0.5cm}
\caption{Heavy $(Q \overline{Q})$ realization via deformation of D3-brane
world-volume. Highly non-BPS configuration
(a) corresponds to two throats located
at the same point on $\Omega_5$. For BPS configuration (b), two throats are
at anti-podal points on $\Omega_5$. By continuous rotation on $\Omega_5$,
(b) can be turned into (a) and vice versa. }
\end{figure}

While an explicit solution describing $Q \overline{Q}$ configuratoin might be
possible, we were not able to find the solution in any closed form starting
from the quantum Born-Infeld action. Therefore,
in this Section, we will calculate the static potential for the non-BPS 
$Q \overline{Q}$ configuratoin with asymptotic approximation. 
Namely, if the separation between the semi-infinite string representing
quark and another representing anti-quark is wide enough, the field 
configuration may be approximately to a good degree by a linear superposition
of two pair of a single string BPS solution with opposite sign choice in
Eq.(\ref{higgsfield}). For $\alpha'$ U field, the approximate configuration 
is given by
\be
{\rm U}({\bf r}) \,\, := \,\, {\rm U}_0 + n \lambda_{\rm IIB} 
\left( {1 \over |{\bf r} + {\bf d}/2|}
+ {1 \over |{\bf r} - {\bf d}/2| } \right),
\label{approx}
\ee
while the electric field is a linear superposition of difference of the
gradients of each term in Eq.(\ref{approx}).
Note that the inflection point of $\alpha'$U field is  around the 
midpoint ${\bf r} = 0$ between $Q$ and $\overline{Q}$.
If we denote the lift of $U$-field at this point as $U_*$, measured 
relative to the asymptotic one ${\rm U}_0$, it is given by 
\be
{\rm U}_* \approx {4 n \over |{\bf d}|}.
\label{dualityrelation}
\ee
Interestingly, short-distance limit (i.e. inter-quark separation $ \vert 
{\bf d} \vert \rightarrow 0$) in the gauge 
theory corresponds to a long-distance limit ($ {\rm U}_* \rightarrow \infty$) 
in the anti-de Sitter supergravity and vice versa.

Let us now estimate the static $Q \overline{Q}$ potential.
If we insert the linear superposition of solutions to the energy functional, 
Eq.(39), there are self-energy contributions of the form precisely as in
the last line in Eq.(39). Subtracting out (or rather renormalizing) these
self-energies, we are left with interaction energy:
\be
V(d) \sim
2 n^2 \int d^3 x \, {1 \over |{\bf r} + {\bf d}/2|^2 } 
{1 \over |{\bf r} - {\bf d}/2|^2 } (\hat {\bf r} + {\hat {\bf d}}/2)
\cdot (\hat {\bf r} - {\hat {\bf d}}/2).   
\ee
The integral is finite and, by dimensional analysis, is equal to
\be
V_{Q \overline{Q}} (d) \hskip0.5cm \sim \hskip0.5cm 
2 n^2 {C_{\rm BI} \over |{\bf d}|},
\label{pota}
\ee
where the coefficient $C_2$ depends on $g_{\rm st}$ and $N$. 
The dimensionless numerical coefficient $C_{\rm BI}$, which depends critically
on $\lambda_{\rm IIB}$ and $N$ through the relation Eq.(42), 
can be calculated, for example, by Feynman parametrization method. 
The interaction potential is indeed Coulomb potential -- inversely proportional
to the separation and proportional to charge-squared. Utilizing the `geometric 
duality' relation Eq.(\ref{dualityrelation}), it is also possible to 
re-express the static potential as:
\be
V_{Q \overline{Q}} ({\rm U}_*) \hskip0.5cm \sim \hskip0.5cm {1 \over 2} n 
{\rm U}_* C_{\rm BI} \, .
\label{potb}
\ee
Recall that ${\rm U}_*$
was a characteristic measure of $\alpha'$U field lift relative to the asymptotic 
value $U_0$ (See figure 3). Since this is caused by bringing $Q$ and 
$\overline{ Q}$
of {\sl same} orientation, the interpretation would be that the static 
$Q \overline{ Q}$ potential is produced by ${\rm U}_*$ 
portion of the string due to the
presence of neighbor non-BPS string. In some sense, the $Q \overline{Q}$ pair
experiences a confining force in $\alpha'$ U direction. The fact that 
Eq.(\ref{potb}) is proportional {\sl linearly} to the charge $n$ is another 
hint to this `dual' interpretation.
The result Eq.(\ref{potb}), however, does not expose the aforementioned
non-analyticity of square-root branch cut type in the previous subsection.
We interpret this provisionally as asssertion that the Born-Infeld theory
is insufficient for full-fledged description of the strong coupling dynamics. 

Now that we have found two distinct $Q \overline{Q}$ configurations, we can 
estimate $Q \overline{Q}$ static potential purely due to Coulomb interaction.
Recall that, for BPS $Q \overline{Q}$ configuration, the Coulomb interaction energy
was cancelled by the $\alpha'$U field gradient energy. On the other hand, for
non-BPS $Q \overline{Q}$ configuration, the two adds up. Thus, by taking an average
of the two, we estimate 
that purely Coulomb potential between static $Q \overline{Q}$ equals to half of 
Eq.(\ref{pota}) or, equivalently, of Eq.(\ref{potb}).

%%%%%%%%%%%%%%%%%%%%%%%%%%%%%%%%%%%%%%%%%%%%%%%%%%%%%%%%%%%%%%%%%%%%%%%%%%%%
\subsection{Heavy Quark Potential in One Dimension}
%%%%%%%%%%%%%%%%%%%%%%%%%%%%%%%%%%%%%%%%%%%%%%%%%%%%%%%%%%%%%%%%%%%%%%%%%%%%
In the previous subsection, we have estimated the $Q \overline{Q}$ static potential
only approximately by linearly superimposing two oppsite sign BPS string 
configurations. To ascertain that this is a reasonable approximation, we
study a simpler but exactly soluble example of $Q \overline{Q}$ potential:
a pair of oppositely oriented fundamental strings hung over two parallel,
widely separated D-strings.  

Consider, as depicted in figure 4, a pair of D-strings of length $L$ along
$x$-directon, whose ends are at fixed position. 
The two fundamental strings of opposite orientation are connected to the
two D-strings and are separated by a distance $d$ in $x$-direction. 
At $\lambda_{\rm IIB} \rightarrow 0$, the fundamental strings obey 
the Polchinski's string boundary conditions and are freely sliding
on the D-string. 
 
\begin{figure}[t]
   \vspace{0cm}
   \epsfysize=8cm
   \epsfxsize=10cm
   \centerline{\epsffile{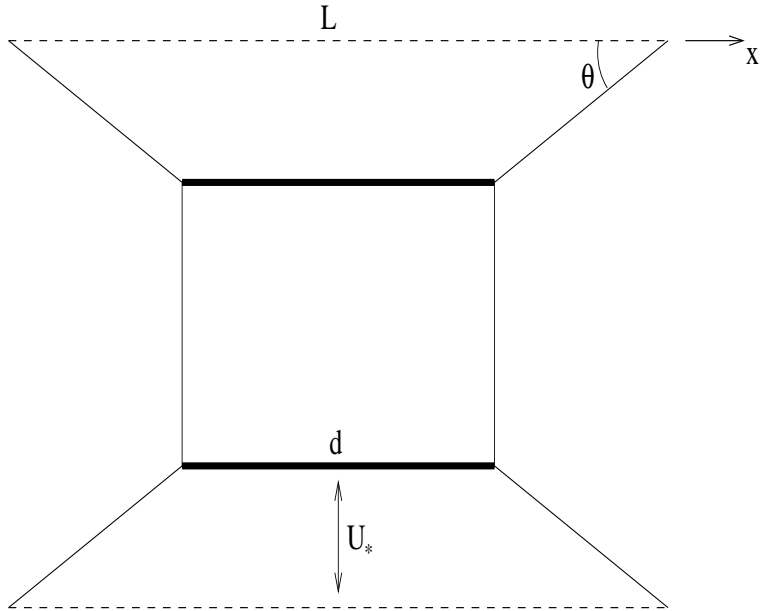}}
\caption{
Non-BPS configuration of quark anti-quark pair on D-string. }
\end{figure}

Once $\lambda_{\rm st}$ is turned on, the string network get deformed
into a new equilibrium configuration. It is intuitively clear what will
happen: the two fundamental strings will attract the two D-strings.
In doing so, length of fundamental strings is shortened. Since the two
funadamental strings are oppositely oriented, they will attract each other
and eventually annihilate. In the weak coupling regime, however, the force
is weak compared to the inertial mass of the fundamental string. We shall 
calculate the potential between them in this weak coupling regime.   

This energy difference is given by:
\bee
V_{Q \overline{Q}} (d) &=& d \left[ \sqrt{{1 \over \lambda^2_{\rm IIB}} + n^2} - 
{1 \over \lambda_{\rm IIB}} \right]
\nonumber \\
&\approx& d \left[ {1 \over 2} n^2 \lambda_{\rm IIB} \right]
.
\label{dstringpot}
\eee
This indeed represent the static$(Q \overline{Q})$ potential. As expected for
Coulomb interaction, the energy is proportional to quark charge-squared
$n^2$. It is also proportional to string coupling $\lambda_{\rm st}
$, which is also proportional to $g^2_{\rm YM}$. 

The potential can be interpreted differently. The four portions of D-strings
between each string junctions and the fixed ends are now all bent by the
same angle $\theta$ relative to the $x$-axis.
From the requirement of tensional force balance at each string triple 
junction one finds easily that
\be
\tan \theta = n \, \lambda_{\rm IIB}.
\ee
Then, simple geometric consideration leads to the relation that
the shortening of the fundamental string denoted by $U_*$ is given by
\be
2{\rm U}_* =  
(L - d) \tan \theta.
\ee
Using these relations to Eq.(\ref{dstringpot})
we now find that 
\be
V_{Q \overline{Q}} ({\rm U}_*) = n \,  {\rm U}_* 
\ee
plus irrelevant bulk contribution.
In this alternative form, it is clear that the static potential energy
originates from the deformation of the string network, which in turn
reduces the length of fundamental strings. 

Note that, in deriving the above results, we have linearly superposed
two triple string junctions, each satisfying BPS conditions 
$E = \pm \nabla_x {\rm U}$ respectively. The linearly superposed configuration 
then breaks supersymmetries completely. Nevertheless, at weak coupling and 
for macroscopically large size, we were able to treat the whole problem 
quasi-statically, thanks to the (almost) infinite inertial mass of the 
fundamental strings. Thus, approximations and results are exactly the same 
as for $(Q \overline{Q})$ on D3-branes.  

%%%%%%%%%%%%%%%%%%%%%%%%%%%%%%%%%%%%%%%%%%%%%%%%%%%%%%%%%%%%%%%%%%%%%%%%%%%%
\subsection{$\theta$-Dependence of Inter-Quark Potential}
%%%%%%%%%%%%%%%%%%%%%%%%%%%%%%%%%%%%%%%%%%%%%%%%%%%%%%%%%%%%%%%%%%%%%%%%%%%%
The $d=4, {\cal N} = 4$ super Yang-Mills theory contains two coupling
parameters $g_{\rm YM}^2$ and $\theta$, the latter being a coefficient
of ${\rm Tr} ( \epsilon^{\mu \nu \alpha \beta} F_{\mu \nu} F_{\alpha \beta} )
/ 32 \pi^2$. From the underlying Type IIB string theory, they arise from 
the string coupling parameter $\lambda_{\rm st}$ and Ramond-Ramond zero-form
potential $C_0$. They combine into a holomorphic coupling parameter
\bee
\tau &=& {\theta \over 2 \pi} + i { 4 \pi \over g^2_{\rm YM} N}
\nonumber \\
&=& C_0 + i {1 \over \lambda_{\rm IIB}}.  
\eee

From the gauge theory point of view, one of the interesting question is
$\theta$-dependence of the static quark potential. Under $d=4$ P and
CP, the former is odd while the latter is even. Thus, the static 
quark potential should be an even function of $\theta$. 
The $\theta$ ranges $(0, 2 \pi)$. Then, the periodicity of $\theta$
(i.e. T-transformation of $SL(2,{\bf Z})$ and invariance of static quark
potential under parity transformation dictate immediately that the
quark potential should be symmetric under $\theta \rightarrow - \theta$
and $\pi - \theta \rightarrow \pi + \theta$. This yields cuspy form of
the potential. Since the whole physics descends from the $SL(2,{\bf Z})$
S-duality, let us make a little calculation in a closely related system:
the triple junction network of $(p,q)$ strings. This system will exhibit
most clearly the very fact that string tension is reduced most at 
$\theta = \pi$. 
That this is so can be seen from replacing $n$ in the previous analysis
by $\theta$-angle rotated dyon case:
\be
n \rightarrow \sqrt{ (n - \theta m)^2 + {m^2 \over \lambda_{\rm IIB}^2} }.
\ee

The whole underlying physics can be understood much clearer from 
the D-string junctions. Consider a $(0,1)$ D-string in the background
of Ramond-Ramond zero-form potential. The Born-Infeld Lagrangian reads
\be
L_{D1} = {T \over \lambda_{\rm IIB}}
\int dx \, \sqrt{ 1 + (\nabla X)^2 - F^2} + C_0 \wedge F 
\ee
Consider a $(1,0)$ fundamental string attached on D-string at location
$x = 0$. The static configuration of the triple string junction is then
found by solving the equation of motion. In $A_1 = 0$ gauge, 
\be
\nabla \left( { -  \nabla A_0  \over \sqrt{1 + (\nabla X_9)^2 - (\nabla
A_0)^2} } - \lambda_{\rm IIB} C_0 \right) = \lambda_{\rm IIB} \delta (x).
\ee
The solution is $X_9 = {\sqrt a} A_0$ for a continuous parameter $a$,
where
\be
{ \nabla A_0 \over \sqrt {1 - (1 - a) (\nabla A_0)^2} }
= \lambda_{\rm st} \theta(x_1) - \lambda_{\rm IIB} C_0.
\ee
Substituting the solution to the Born-Infeld Lagrangian, we find the
string tension of D-string:
\bee
T_D = \left\{ \begin{array}{cc}
\sqrt {{1 \over \lambda_{\rm IIB}^2} + (1 - C_0)} & \hskip1cm  x_1 > 0
\\
\sqrt{{1 \over \lambda_{\rm IIB}^2} + C_0^2} & \hskip1cm x_1 < 0.
\end{array}
\right.
\nonumber
\eee
Clearly, tension of the $(1,1)$ string (on which electric field is
turned on) attains the minimum when $C_0 = 1/2$, viz. $\theta = \pi$. 
Moreover, in this case, the D-string bends symmetrically around the
junction point $x_1 = 0$, reflecting the fact that P and CP symmetries
are restored at $\theta = \pi$. 
%%%%%%%%%%%%%%%%%%%%%%%%%%%%%%%%%%%%%%%%%%%%%%%%%%%%%%%%%%%%%%%%%%%%%%%%%%%%
\section{Further Considerations}
%%%%%%%%%%%%%%%%%%%%%%%%%%%%%%%%%%%%%%%%%%%%%%%%%%%%%%%%%%%%%%%%%%%%%%%%%%%%
In this Section, we take up further the present results and speculate 
two issues that might be worthy of further study.

%%%%%%%%%%%%%%%%%%%%%%%%%%%%%%%%%%%%%%%%%%%%%%%%%%%%%%%%%%%%%%%%%%%%%%%%%%%%
\subsection{Dynamical Realization of Large-N Loop Equation}
%%%%%%%%%%%%%%%%%%%%%%%%%%%%%%%%%%%%%%%%%%%%%%%%%%%%%%%%%%%%%%%%%%%%%%%%%%%%
It is well-known that the Wilson loop
\be
W[X] = \exp \oint_C ds {\dot X}^M A_M(X(s))
\ee
satisfies the classical identity
\be
\int_0^\epsilon d \sigma {\delta^2 \over \delta X_M(+\epsilon)
\delta X_M(\epsilon)} W[X]
= \nabla^M F_{MN} (X(0)) {\dot X}^N (0) W [X].
\ee
Physically, the equation can be interpreted as a variation of the Wilson
loop as the area enclosed is slightly deformed.

More recently, based on dual description of large N gauge theory in 1+1
dimensions in terms of near-critical electric field on a D-string,
Verlinde~\cite{verlinde} has shown that Wilson loop equation follows as 
the conformal Ward identity on the string world-sheet. 
Immediate question that arise is, relying on the $SO(4,2)$ 
conformal invariance of large $N$ super Yang-Mills gauge theory,
whether one can extend the Verlinde's result and derive the large $N$ loop
equation. In what follows, we would like to present rather heuristic
arguments why and how conformal invariance might play some role in this 
direction.

Classically large $N$ loop equation asserts invariance of the Wilson loop
average under small variation of the area enclosed by the loop. Let us now
restrict ourselves to timelike Wilson loops and apply a small deformation 
of the contour $C$. As the contour $C$ of timelike Wilson loop represents 
straight world-line of heavy quark and anti-quark pair, the adiabatic
local deformation of the Wilson loop may be interpreted as a result of 
acceleration of initially static quark $Q$ and subsequent deceleration
back to the original static quark worldline during a small time interval. 
This is depicted in figure 5(a). Normally, such acceleration and deceleration
requires turning on and off some adiabatic electric field in the region 
near the quark $Q$ trajectory ( the shaded region of figure 5(a)).  

\begin{figure}[t]
   \vspace{0cm}
   \epsfysize=7cm
   \epsfxsize=13cm
   \centerline{\epsffile{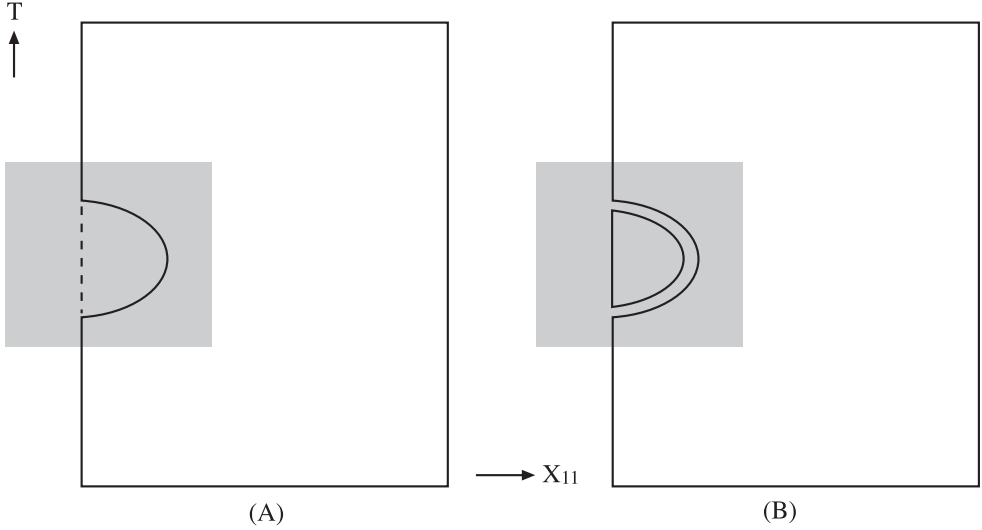}}
\vspace{0.5cm}
\caption{Conformal transformation cause local recoil of timelike loop. 
Back tracking at large $N$ is equivalent to pair creation process.}
\end{figure}

However, for conformally invariant Yang-Mills theory, there is an amusing
possibility that accelerating (decelerating) charge configuration can be 
achieved via conformal transformation {\sl without} background electric field. 
Recall that, in Lorentz invariant theory, it is always possible that a static
configuration can be brought into a uniformly boosted configuration by an 
application of Lorentz transformation. What conformally invariant theory does
is more than that and can even relate, for example, uniformly accelerated
(decelerated) configuration by conformal transformation to a static (or
uniformly boosted) configuration. Indeed, if we apply a special $SO(4,2)$
conformal transformation of an inversion with a translation by $a^\mu$
followed by another inversion,
\be
x^\mu \rightarrow x^{\mu'} = {x^\mu + a^\mu x^2 \over
1 + 2 a \cdot x + a^2 x^2}.
\ee
If we set $a^\mu = (0 , - {1\over 2} {\bf a})$, we obtain
\be
t' = {t \over 1 - {\bf z} \cdot {\bf x} + {1 \over 4} {\bf a}^2 (
{\bf x}^2 - t^2)}, \hskip0.3cm
{\bf x}' = {{\bf x} + {1 \over 2} {\bf z} (t^2 - {\bf x}^2)
\over 1 - {\bf a} \cdot {\bf x} + {1\over 4} {\bf z}^2 ({\bf x}^2 - t^2)}.
\ee
Thus, the original trajectory of static configuration at ${\bf x} = 0$ is
now transformed into
\be
t_* = {t \over 1 - {1 \over 4} {\bf a}^2 t^2}, \hskip0.5cm
{\bf x}_* = { {1 \over 2} {\bf a} t^2 \over 1 - {1 \over 4} {\bf a}^2 t^2},
\ee
which, for $|t| < 2/|{\bf a}|$,
represents the coordinates of a configuration with constant acceleration
${\bf a}$ passing through the origin ${\bf x}_* = 0$ at $t_* = 0$.

Thus, if one performs instantaneous special conformal transformations on
a finite interval along the heavy quark $Q$ trajectory, then it would be
indeed possible to show that a timelike Wilson loop is equivalent to a 
deformed Wilson loop (by the conformal transformation, however, only 
timelike deformations can be realized). Since the anti-podally oriented
$Q \overline{Q}$ pair is a BPS state, it might even be possible to generate 
a four-quark (of which two are virtual BPS states) intermediate 
state by a variant of the conformation transformation, as depicted in 
figure 5(b). 
Details of this issue will be reported elsewhere~\cite{bakrey}.

%%%%%%%%%%%%%%%%%%%%%%%%%%%%%%%%%%%%%%%%%%%%%%%%%%%%%%%%%%%%%%%%%%%%%%%%%%%%
\subsection{Multi-Prong Strings}
%%%%%%%%%%%%%%%%%%%%%%%%%%%%%%%%%%%%%%%%%%%%%%%%%%%%%%%%%%%%%%%%%%%%%%%%%%%%
Moving a step further, can we manufacture a static configuration that may 
be an analog of {\sl baryon} in QCD out of Type IIB strings? For the 
gauge group $SU(N)$, the baryon is a gauge singlet configuration obeying 
$N$-ality. Clearly, we need to look for a string configuration that can be
interpreted as a $N$-quark state on the D3-brane world-volume. Recently,
utilizing triple BPS string junction~\cite{reyyee, triple},
such a configuration has been identified~\cite{bergman}: $N$-pronged string
junction interconnecting $N$ D3-branes. For example, for gauge group $SU(3)$ 
realized by three D3-branes, multi-monopole configuration that may be interpreted as the static baryon is a triple string junction as depicted in figure 6.
The $N$-pronged string junction is a natural generalization of this, as can be
checked from counting of multi-monopole states and comparison with the
$(p,q)$ charges of the Type IIB string theory. 

\begin{figure}[t]
   \vspace{0cm}
   \epsfysize=7cm
   \epsfxsize=13cm
   \centerline{\epsffile{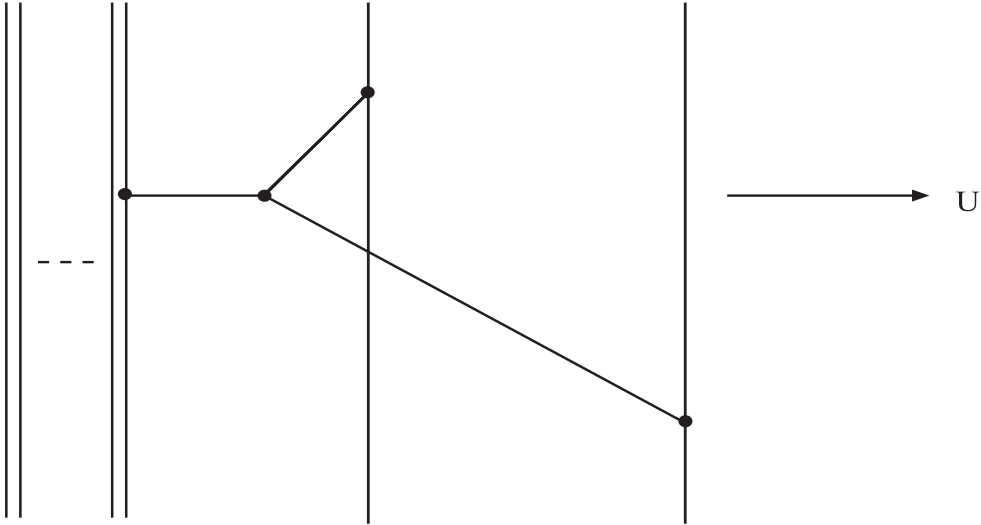}}
\vspace{0.5cm}
\caption{Macroscopic string as a BPS soliton on D3-brane worldvolume.
Large-N corrections induced by branes at $U = 0$ in general gives rise
to corrections to the shape and low-energy dynamics of the D3-brane.}
\end{figure}

The $N$-pronged string junction also exhibits dynamics of marginal stability
as we move around the D3-branes on which each prongs are 
attached~\cite{bergman}. Adapted to the present context, for example, in
figure 6 situation, this implies that as the triple junction point is moved
around by moving the position of the two outward D3-branes as well as their
$\Omega_5$ angular coordinates, the triple string junction will decay once
the inner prong becomes shorter below the curve of marginal stability.
The final configuration is easily seen to be a pair of macroscopic
strings, each one connecting to  the two outer D3-branes separately.

%%%%%%%%%%%%%%%%%%%%%%%%%%%%%%%%%%%%%%%%%%%%%%%%%%%%%%%%%%%%%%%%%%%%%%%%%%%%
\subsection{Quarks and $(Q \overline{Q})$ at Finite-Temperature}
%%%%%%%%%%%%%%%%%%%%%%%%%%%%%%%%%%%%%%%%%%%%%%%%%%%%%%%%%%%%%%%%%%%%%%%%%%%%
So far, our focus has been, via the AdS-CFT correspondence, the holographic 
description of strongly coupled $N=4$ super Yang-Mills theory at zero 
temperature. The AdS-CFT correspondence, however, is not only for the super
Yang-Mills theory at zero temperature but also extendible for the theory
at finite temperature. Is it then possible to understand finite-temperature
physics of quark dynamics and static quark potential, again, from the AdS-CFT
correspondence? We will relegate the detailed analysis to a separate work
\cite{rty}, and, in this subsection, summarize what is known from the super
Yang-Mills theory side and propose the set-up for holographic description.

At a finite critical temperature $T = T_c$ , pure $SU(N)$ gauge theory 
exhibits a deconfinement phase transition.
The relevant order parameter is the Wilson-Polyakov loop:
\be
P({\bf x}) = {1 \over N} {\rm Tr} {\cal P} \exp \left( i \int_0^{1 \over T} 
A_0 ({\bf x}) dt \right).
\label{polyakov}
\ee
Below the critical temperature $T < T_c$, $\langle P \rangle = 0$ and
QCD confines. Above  $T > T_c$, $\langle P \rangle$ is nonzero and takes
values in ${\bf Z}_N$, the center group of $SU(N)$.
Likewise, the two-point correlation of parallel Wilson-Polyakov loops 
\bea
\Gamma({\bf d}, T) \quad \equiv \quad \langle P^\dagger({\bf 0}) 
\,  P ({\bf d}) \rangle_T
\quad = \quad e^{- {\cal F}({\bf d},T)/T } \,\, \approx 
\,\, e^{- V_{\rm Q {\overline Q}} ({\bf d}, T)/T }
\eea
measures the static potential at finite temperature between quark and
anti-quark separated by a distance $d$.

At sufficiently high temperature, thermal excitations produce a plasma 
of quarks and gluons and gives rise to Debye mass $m_{\rm E} \approx
g_{\rm eff} T$ (which is responsible for screening color electric flux) and magnetic mass 
$m_{\rm M} \approx g^2_{\rm eff} T$ (which corresponds to the glueball mass
gap in the confining three-dimensional pure gauge theory). Their effects 
are captured by the asymptotic behavior of the heavy quark 
potential:
\bee
\begin{array}{llll}
V_{\rm Q {\overline Q}} ({\bf d}, T)
\,\,\, \approx & - \, C_{\rm E} \, 
{1 \over |{\bf d}|^2} \,e^{- 2 \, m_{\rm E} |{\bf d}|} 
\,\, + \cdots & \qquad \qquad \quad & C_{\rm E} = {\cal O}(g^4_{\rm YM})\\
&&& \\
 & -  C_{\rm M} \, 
{1 \over |{\bf d}|} \, e^{- m_{\rm M} |{\bf d}|}
\,\, + \cdots &\qquad \qquad \quad & C_{\rm M} = {\cal O}(g^{12}_{\rm YM})
.  \end{array}
\label{debye}
\eee

At finite temperature, it is known that the large
$N$ and strong coupling limit of $d=4, {\cal N}=4$ supersymmetric gauge 
theory is dual to the near-horizon geometry of near extremal 
D3-branes in Type IIB string theory.
The latter is given by a Schwarzschild-anti-de Sitter Type IIB supergravity
compactification:
\be
ds^2 = \alpha' \left[
{1 \over \sqrt G} \left( - H dt^2 + d {\bf x}_{\parallel}^2 \right)
+ \sqrt{G} \left( {1 \over H} d {\rm U}^2 + {\rm U}^2 d {\bf \Omega}_5^2
\right) \, \right]
\label{sugra}
\ee
where
\bee
G &\equiv& {g^2_{\rm eff} \over {\rm U}^4} 
\nonumber \\
H &\equiv& 1 - { {\rm U}_0^4 \over {\rm U}^4 } 
\hskip1cm \left({\rm U}_0^4 = {2^7 \pi^4 \over 3 } g^4_{\rm eff} \, 
{\mu \over N^2}
\right) \, .
\label{harmonicftn}
\eee
The parameter $\mu$ is interpreted as the free energy density on the near
extremal D3-brane, hence, $\mu = (4 \pi^2 /45) N^2 T^4$. In the 
field theory limit $\alpha' \rightarrow 0$, $\mu$ remains finite. In turn, 
the proper energy $ E_{\rm sugra} = \sqrt{g_{\rm eff} / \alpha'} \mu / {\rm U}$
and the dual description in terms of modes propagating in the above
supergravity background is expected to be a good approximation.  

Hence, the question is whether the Debye screening of the static quark
potential Eq.(\ref{debye}), or any strong coupling modification thereof, can 
be understood from the holographic description in the background 
Eq.(\ref{sugra}). In \cite{rty}, we were able to reproduce a result 
qualitatively in agreement with Eq.(\ref{debye}). The strong coupling 
effect again shows up throgh the non-analytic dependence of the potential
to the `t Hooft coupling parameter, exactly the same as for zero-temperature
static potential. In \cite{kr}, we have also found a result indicating that
the finite-temperature free energy of ${\cal N}=4$ super Yang-Mills theory
interpolates smoothly with the `t Hooft coupling parameter, barring a possible
phase transition between the weak coupling and the strong coupling regimes. 
%%%%%%%%%%%%%%%%%%%%%%%%%%%%%%%%%%%%%%%%%%%%%%%%%%%%%%%%%%%%%%%%%%%%%%%%%%%%
\section{Discussion}
%%%%%%%%%%%%%%%%%%%%%%%%%%%%%%%%%%%%%%%%%%%%%%%%%%%%%%%%%%%%%%%%%%%%%%%%%%%%
In this paper, we have explored some aspects of the proposed relation between 
$d=4, {\cal N}=4$
supersymmetric gauge theory and maximal supergravity on $AdS_5 \times S_5$
using the Type IIB $(p,q)$ strings as probes. From the point of view of D3
brane and gauge theory thereof, semi-infinite strings attached on it are
natural realization of quarks and anti-quarks. Whether a given configuration
involving quarks and anti-quarks is a BPS configuration does depend
on relative orientation among the strings (parametrized by angular coordinates 
on $S_5$). The physics we have explored, however, did not rely much on it
since the quarks and anti-quarks have infinte inertia mass and are nominally 
stable. 

The results we have obtained may be summarized as follows. For a single
quark $Q$ (or anti-quark $\overline{Q}$) 
BPS configuration, near-extremal excitation  corresponds
to fluctuation of the fundamental string. We have found that the governing
equations and boundary conditions do match precisely between the large-$N$
gauge theory and the anti-de Sitter supergravity sides. In due course, we
have clarified the emergence of Polchinski's D-brane boundary condition (
Dirichlet for perpendicular and Neumann for parallel directions) 
as the limit $\lambda_{\rm IIB} \rightarrow 0$ is taken.  
For non-BPS $Q \overline{Q}$ pair configuration, we first have studied 
inter-quark
potential and again have found an agreement between the gauge theory and
the anti-de Sitter supergravity results. Measured in units of Higgs 
expectation value, the potential exhibits linear potential that allows an
interpretation of confinement. Because the theory has no mass gap generated
by dimensional transmutation, the fact that string tension is measured in
units of Higgs expectation value may not be so surprising. We have also
explored $\theta$-dependence of the static quark potential by turning on
a constant Ramond-Ramond 0-form potential. The $SL(2,{\bf Z})$ S-duality
of underlying Type IIB string theory implies immediately that the static
quark potential exhibits cusp behavior at $\theta = \pi$. The potential
strength is the weakest at this point and hints a possible realization of
deconfinement transition at $\theta = \pi$.   
We also discussed qualitatively two related issues. Via conformal invariance
we have pointed out that a static quark configuration can be transformed
into an accelerating (or decelerating) configuration. Viewed this as
a physical realization of deforming the Wilson loop, we have conjectured
that it is this conformal invariance that allows to prove the large $N$
Wilson loop equation for a conformally invariant super Yang-Mills theory. 
We also argued that an analog of static baryons $(Q \cdots Q)$ in QCD are 
represented by multi-prong string junctions. 

We think the results in the present paper may be of some help eventually
in understanding dynamical issues in the large $N$ limit of 
superconformal gauge theories. For one thing, it would be very interesting
 to understand dynamical light or massless quarks and physical excitation 
spectra. While we have indicated that qualitative picture of the excitation 
spectrum as conjectured by Maldacena would follow from near-extremal 
excitation of fundamental strings themselves, a definitive
answer awaits for a full-fledged study.

\vskip0.5cm
SJR thanks D. Bak, C.G. Callan, I. Klebanov, J.M. Maldacena, 
and H. Verlinde and other participants of Duality '98 Program at the Institute 
for Theoretical Physics at Santa Barbara for useful discussions. 
SJR is grateful to the organizers M.R. Douglas, W. Lerche and H. Ooguri for 
warm hospitality. 

%%%%%%%%%%%%%%%%%%%%%%%%%%%%%%%%%%%%%%%%%%%%%%%%%%%%%%%%%%%%%%%%%%%

\end{document}